\documentclass[12pt]{article}
\pdfoutput=1

\usepackage{cite}
\usepackage{booktabs}
\usepackage[english]{babel}
\usepackage{amsmath,amssymb,amsbsy,amstext, amsthm, simplewick}
\usepackage{hyperref}
\usepackage{graphicx}
\usepackage{amsfonts}
\usepackage{amssymb}
\usepackage[small]{caption}
\usepackage{upgreek}
\usepackage[svgnames,dvipsnames,x11names,table]{xcolor}
\usepackage{multirow}
\usepackage{geometry}
\usepackage[hang,flushmargin]{footmisc}
\usepackage{bm}
\usepackage{braket}
\usepackage{subcaption}
\usepackage{mathtools}
\usepackage{setspace}
\usepackage{cleveref}
\usepackage{comment}
\usepackage{scalerel}
\usepackage[normalem]{ulem}
\usepackage{slashed}
\usepackage{enumitem}
\usepackage{dsfont}
\usepackage{tikz}
\usepackage{colortbl}
\usepackage[table]{xcolor}
\usetikzlibrary{decorations.markings}
\usetikzlibrary{shapes.misc}

\definecolor{colorC0}{HTML}{1f77b4}
\definecolor{colorC1}{HTML}{ff7f0e}
\definecolor{colorC2}{HTML}{2ca02c}

\newcolumntype{d}{>{\columncolor{colorC0!10}}c}
\newcolumntype{e}{>{\columncolor{colorC1!10}}c}
\newcolumntype{f}{>{\columncolor{colorC2!10}}c}

\makeatletter
\g@addto@macro\bfseries{\boldmath}
\makeatother

\hypersetup{
    colorlinks=true,
    linkcolor={red!50!black},
    citecolor={blue!50!black},
    urlcolor={blue!80!black}
}

\usepackage{colortbl}

\makeatletter
\newlength{\apb@width}
\newcommand{\autoparbox}[2][c]{\settowidth{\apb@width}{#2}\parbox[#1]{\apb@width}{#2}}

\makeatother

\definecolor{lightgray}{gray}{0.9}

\usepackage[framemethod=default]{mdframed}
\newmdenv[skipabove=7pt,
skipbelow=7pt,
rightline=false,
leftline=false,
topline=false,
bottomline=false,
backgroundcolor=gray!10,
linecolor=gray,
innerleftmargin=5pt,
innerrightmargin=5pt,
innertopmargin=5pt,
innerbottommargin=5pt,
leftmargin=0cm,
rightmargin=0cm,
linewidth=4pt]{eBox}

\usepackage[most]{tcolorbox}
\tcbset{colback=white, colframe=black,
        highlight math style= {enhanced, %<-- needed for the ?remember? options
            colframe=red,colback=red!10!white,boxsep=0pt}
        }
\definecolor{light-gray}{gray}{0.95}

\crefname{table}{Table}{Tables}
\crefname{equation}{Eq.}{Eqs.}
\crefname{appendix}{App.}{Apps.}
\crefname{section}{Sec.}{Secs.}
\crefname{figure}{Fig.}{Figs.}

\numberwithin{equation}{section}

\def\beq{\begin{equation}}
\def\eeq{\end{equation}}

\def\bea{\begin{eqnarray}}
\def\eea{\end{eqnarray}}

\def\Neff{N_{\rm eff}}
\def\beq{\begin{equation}}
\def\eeq{\end{equation}}
\def\bea{\begin{eqnarray}}
\def\eea{\end{eqnarray}}

\def\tnl{\tau_{\rm NL}}

\def\k{{\vec{\scaleto{k}{7pt}}}}

\def\x{{\vec x}}

\DeclareRobustCommand{\SkipTocEntry}[4]{}

\definecolor{colorTC}{rgb}{.2,.7,.2}

\definecolor{amethyst}{rgb}{0.6, 0.4, 0.8}

\definecolor{acolor}{rgb}{0.4, 0.2, 0.4}

\definecolor{blue3}{RGB}{31, 119, 180}
\definecolor{red3}{RGB}{	214, 39, 40}
\definecolor{orange3}{RGB}{255, 127, 14}
\definecolor{green3}{RGB}{44, 160, 44}

\begin{document}

\begin{titlepage}
\setcounter{page}{1} \baselineskip=15.5pt
\thispagestyle{empty}
$\quad$
\vskip 70 pt

\begin{center}

{\fontsize{20.74}{24} \bf No $\nu$s is Good News}
\end{center}

\vskip 20pt
\begin{center}
\noindent
{\fontsize{12}{18}\selectfont Nathaniel Craig$^{1,2}$, Daniel Green$^3$, Joel Meyers$^4$, and Surjeet Rajendran$^5$}
\end{center}

\begin{center}
\vskip 4pt
\textit{$^1${\small Department of Physics, University of California, Santa Barbara, CA 93106, USA}\\
\textit{$^2${\small Kavli Institute for Theoretical Physics, Santa Barbara, CA 93106, USA}}\\
$^3${\small Department of Physics, University of California, San Diego,  La Jolla, CA 92093, USA}\\
$^4${\small Department of Physics, Southern Methodist University, Dallas, TX 75275, USA}\\
 $^5${\small Department of Physics \& Astronomy, The Johns Hopkins University, \\
 Baltimore, MD  21218, USA}}
\end{center}

%=========================================
\vspace{0.4cm}
 \begin{center}{\bf Abstract}
 \end{center}

\noindent
The baryon acoustic oscillation (BAO) analysis from the first year of data from the Dark Energy Spectroscopic Instrument (DESI), when combined with data from the cosmic microwave background (CMB), has placed an upper-limit on the sum of neutrino masses, $\sum m_\nu < 70$ meV (95\%). In addition to excluding the minimum sum associated with the inverted hierarchy, the posterior is peaked at $\sum m_\nu = 0$ and is close to excluding even the minumum sum, 58 meV at 2$\sigma$. In this paper, we explore the implications of this data for cosmology and particle physics. The sum of neutrino mass is determined in cosmology from the suppression of clustering in the late universe. Allowing the clustering to be enhanced, we extended the DESI analysis to $\sum m_\nu < 0$ and find $\sum m_\nu = - 160 \pm 90$ meV (68\%), and that the suppression of power from the minimum sum of neutrino masses is excluded at $99\%$ confidence. We show this preference for negative masses makes it challenging to explain the result by a shift of cosmic parameters, such as the optical depth or matter density. We then show how a result of $\sum m_\nu =0$ could arise from new physics in the neutrino sector, including decay, cooling, and/or time-dependent masses. These models are consistent with current observations but imply new physics that is accessible in a wide range of experiments. In addition, we discuss how an apparent signal with $\sum m_\nu < 0$ can arise from new long range forces in the dark sector or from a primordial trispectrum that resembles the signal of CMB lensing.

\end{titlepage}
\setcounter{page}{2}

\restoregeometry

\begin{spacing}{1.2}
\newpage
\setcounter{tocdepth}{2}
\tableofcontents
\end{spacing}

\setstretch{1.1}
\newpage

\section{Introduction}

The cosmological measurement of the sum of neutrino masses, $\sum m_\nu$, is one of the most anticipated results from the coming generation of cosmic surveys~\cite{Lesgourgues:2006nd,TopicalConvenersKNAbazajianJECarlstromATLee:2013bxd,Dvorkin:2019jgs}. From the measurement of neutrino flavor oscillations~\cite{ParticleDataGroup:2020ssz}, which precisely determine the mass-squared splittings between neutrino mass eigenstates, it can be inferred that the sum of neutrino masses is necessarily greater than 58~meV. This provides a concrete prediction within the standard cosmological model that should be measurable (or detectable) with planned observations~\cite{Font-Ribera:2013rwa,CMB-S4:2016ple}.

The Dark Energy Spectroscopic Instrument (DESI)~\cite{DESI:2016fyo} is expected to provide the necessary increase in sensitivity to $\sum m_\nu$ to measure the minimum sum at $2$ to $3\sigma$~\cite{Font-Ribera:2013rwa,CMB-S4:2016ple}. Cosmological measurements of neutrino mass rely on the measurement of the clustering of matter on scales smaller than the free-streaming length of neutrinos~\cite{Lesgourgues:2006nd}.  A universe containing massive neutrinos will exhibit suppressed matter clustering compared to a universe with only massless neutrinos.  This measurement can be achieved by combining observations of the cosmic microwave background (CMB) with the measurement of the baryon acoustic oscillations (BAO).  The amplitude of clustering can be determined from the measurement of the CMB lensing power spectrum, and this amplitude is compared to what would be expected in a universe with only massless neutrinos~\cite{Kaplinghat:2003bh}.  In the absence of massive neutrinos, the amplitude of matter clustering is determined by the matter density and the primordial amplitude of scalar fluctuations.  Measurements of the CMB angular power spectra allow for a determination of the primordial fluctuation amplitude.  BAO measurements are needed to measure the abundance of non-relativistic matter to sufficient accuracy to isolate the effect neutrino mass~\cite{Pan:2015bgi}.

The release of the first year BAO analysis with DESI~\cite{DESI:2024mwx}, combined with data from the CMB (Planck 2018~\cite{Aghanim:2019ame,Carron:2022eyg} and ACT DR6 lensing~\cite{ACT:2023dou,ACT:2023kun}), showed a remarkable upper-limit on $\sum m_\nu$, reaching
\beq
\sum m_{\nu} < 70 \, {\rm meV} \, (95\%) \ .
\eeq
This is consistent with an earlier constraint from (e)BOSS of $\sum m_\nu < 82$ meV~\cite{Brieden:2022lsd} using CMB+BAO+Shape parameters (see also~\cite{Palanque-Delabrouille:2019iyz}). The DESI result is sufficient to exclude the minimum mass for an inverted neutrino mass hierarchy, 100 meV, at $\sim 3\sigma$.  However, what is also noteworthy is that the posterior peaks at $\sum m_\nu =0$ and is very close to putting 58 meV in tension with observations. 

\begin{figure}[t!]
	\centering
	\includegraphics[width=0.65\textwidth]{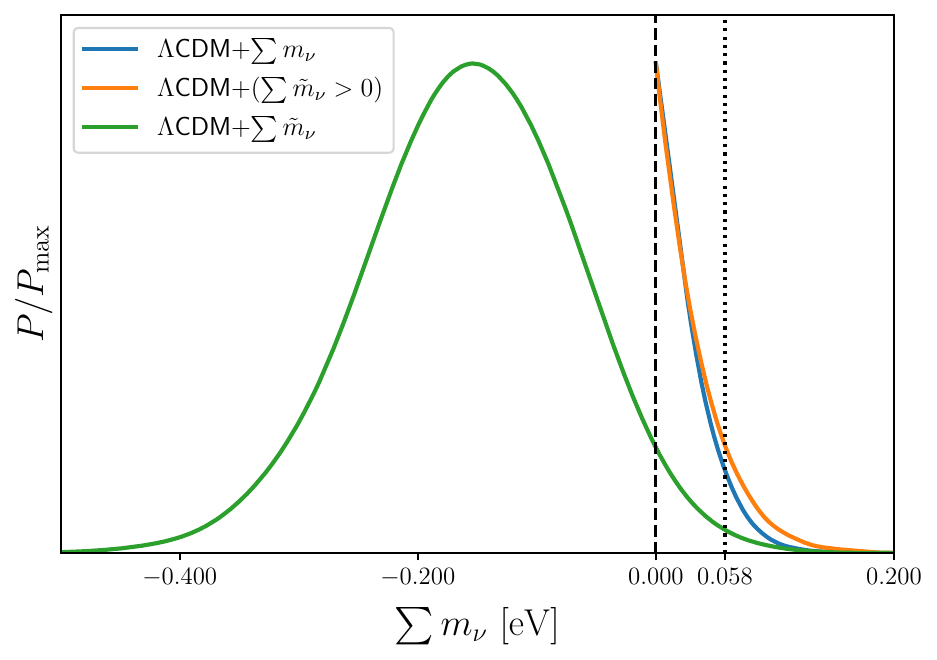}
	\caption{Posterior of neutrino mass in eV inferred from Planck + ACT Lensing + DESI data.  The blue line shows constraints on a model with a physical neutrino mass, the orange line shows constraints where the neutrino mass is parametrized as an effect on the CMB lensing power spectrum and restricted to be positive, and the green line shows constraints on a parametrized neutrino mass that is allowed to be negative.  The best fit for the parametrized neutrino mass is $\sum \tilde{m}_\nu = -160$~meV, and the minimal neutrino mass of 58~meV is disfavored at 3$\sigma$.  For details about the parametrization of negative neutrino mass and the data sets used, see Section~\ref{sec:Negative_mnu}.}
	\label{fig:mnu_constraints}
\end{figure}

In this paper, we will explore the current constraints on $\sum m_\nu$ and what an exclusion of $\sum m_\nu = 58$ meV would mean for cosmology and particle physics. First, we will examine the current measurement and how it depends on different types of surveys. One particularly noteworthy aspect of the DESI measurement is that it appears to favor $\sum m_\nu < 0$, though that region of parameter space was excluded from the DESI analysis by imposing a prior that $\sum m_\nu$ is positive. Although negative neutrino masses are unphysical, a preference in the data for $\sum m_\nu < 0$ may simply reflect an excess of clustering in the late universe, rather than a deficit caused by free streaming neutrinos. We use this idea to define a neutrino mass, $\sum \tilde m_\nu$, that is allowed to be negative and perform the same analysis as DESI without the positive mass prior. We find that that data does prefer negative mass, $\sum \tilde m_\nu = - 160 \pm 90$ meV (68\%), and corresponds to a $3\sigma$ exclusion of the minimum neutrino mass. The full posterior is shown in Figure~\ref{fig:mnu_constraints}.

The preference of the current measurement for negative $\sum m_\nu$ is particularly important as it affects the bias in the measurement of cosmic parameters, particularly the optical depth, $\tau$, that would be required to explain the current limits. For $\ell > 30$, the CMB is only sensitive to the combination $A_s e^{-2\tau}$, where $A_s$ is the amplitude of primordial scalar fluctuations. The determination of $\tau$ is therefore essential for determining $A_s$ and suppression of power a late times, but requires (challenging) large angular scale measurements of the CMB. It is plausible that $\sum m_\nu=0$ could be explained by a statistical or systematic shift in $\tau$, but it is far more challenging to explain $\sum m_\nu = -160$ meV in this way.

%%%%%%%%%%%%%%%%%
%%%%%%%%%%%%%%%%%%

An absence of the neutrino mass signal, while forbidden in the Standard Model (plus neutrino masses), could be a natural consequence of a wide variety of beyond the Standard Model (BSM) scenarios. The most straightforward mechanisms to eliminate the signal would be to eliminate the SM neutrinos via decay (or annihilation), cool the neutrinos so that they behave like dark matter, or change their mass over cosmological history. Simple models for all three scenarios can be derived from new interactions in the neutrino sector that are weakly constrained by experiments. On the other hand, the CMB does provide stringent constraints on the parameter space of these models, as measurements of $\Neff$ are in good agreement with the expected temperature~\cite{Planck:2018vyg} and free-streaming~\cite{Follin:2015hya,Baumann:2015rya,Baumann:2019keh} of the cosmic neutrino background (C$\nu$B). Nevertheless, there have been hints of neutrino interactions~\cite{Cyr-Racine:2013jua,Lancaster:2017ksf,He:2023oke,Camarena:2023cku,Camarena:2024zck} in cosmic data that may also point to new physics of this kind.

Negative neutrino masses, $\sum m_\nu < 0$, are representative of enhanced clustering of matter, rather than any physical property of the neutrinos themselves. This kind of enhanced clustering can be achieved by changing the long range forces that act on matter. We discuss one simple mechanism, which is to introduce a new scalar force that acts only on the dark matter. Such forces are more weakly constrained than fifth forces acting on SM particles and thus could explain our signal without being in tension with other constraints. Alternatively, a CMB lensing measurement with $\sum m_\nu < 0$ points to a larger than expected CMB trispectrum, which could result from a non-zero primordial trispectrum. These scenarios will all be testable with current and/or future cosmic data.

This paper is organized as follows: In Section~\ref{sec:measure}, we review the measurement of $\sum m_\nu$ and extend the analysis to negative masses. We discuss what shifts in cosmic data would be required to make these measurements consistent with conventional neutrino physics. In Section~\ref{sec:zero}, we present models that could explain $\sum m_\nu =0$ with new physics in the neutrino sector. In Section~\ref{sec:negative}, we present models that could explain a cosmological inference of negative neutrino masses. We conclude in Section~\ref{sec:conclusions}. Appendix~\ref{app:A}, we review the physics origin of the suppression of structure due to massive neutrinos.

%%%%%%%%%%%%%%%%%%%%%%%%%%%%%%%%%%%%%%%
\section{Neutrino Mass and DESI}\label{sec:measure}
%%%%%%%%%%%%%%%%%%%%%%%%%%%%%%%%%%%%%%%

\subsection{How Neutrino Mass is Measured}

In order to understand what an apparent measurement of $\sum m_\nu = 0$ would mean, we first need to review exactly what measurements allow us to infer $\sum m_\nu$ (see also~\cite{Green:2021xzn,Green:2022bre} for review). We will assume that $\sum m_\nu \approx 60$~meV, as this is the minimum sum consistent with neutrino oscillation experiments and is therefore the minimum value that would need to be excluded in order to favor $\sum m_\nu = 0$.

Cosmic neutrinos are relativistic in the early universe, but become non-relativistic when their propagation speed, $c_\nu$, drops well below the speed of light. In a $\Lambda$CDM~+~$m_\nu$ cosmology, the typical neutrino speed is given by
\beq
    c_\nu = \frac{\langle p_\nu \rangle}{m_\nu} = \frac{3 T_\nu}{m_\nu} \approx 1.0 \times 10^{-2} \left(\frac{50 \, {\rm meV}}{ m_\nu}\right) \, (1+z)   \ ,
\eeq
where we have set $c=1$. As a result, the redshift where the heaviest neutrino becomes non-relativistic is $z_\nu \approx 100$. For $z< z_\nu$, the energy density of neutrinos redshifts like non-relativistic matter so that 
\beq
\Omega_m = \Omega_{c}+ \Omega_b + \Omega_\nu \ .
\eeq
However, the neutrinos are still sufficiently hot that they do not cluster on scales below their effective Jeans scale. In terms of wavenumber, this free-streaming scale is given by
\beq
    k_{\rm fs} = \sqrt{\frac{3}{2}}\frac{aH}{c_\nu} =  0.04 \, h \, {\rm Mpc}^{-1} \times \frac{1}{1+z} \, \left(\frac{\sum m_\nu}{58 \, {\rm meV}}\right) \, . 
\eeq
Because neutrinos don't cluster, the amplitude of clustering of matter, defined by the matter power spectrum
\beq
    P(k) = \langle \delta_m (\k)\delta_m(\k') \rangle' \ ,
\eeq
is suppressed on scales smaller than the neutrino free-streaming scale $k \gg k_{\rm fs}$
\beq
    P^{(\sum m_\nu)} (k\gg k_{\rm fs}, z ) \approx  \left(1 - 2 f_\nu  -\frac{6 }{5} f_\nu\log \frac{1+z_\nu}{1+z} \right) P^{(\sum m_\nu =0)}(k\gg k_{\rm fs}, z ) \ ,
\eeq
where $f_\nu = \Omega_{\nu} /\Omega_m$ is the fraction of non-relativistic matter in the form of neutrinos, $\delta_m\equiv \delta\rho_m/\bar{\rho}_m$ is the density contrast of non-relativistic matter, and the prime on the correlation function means that the delta function has been omitted. The suppression in this formula is the result of two distinct physical effects (see Appendix~\ref{app:A} for a derivation). The first term, $- 2f_\nu$, reflects the reduced fraction of matter that is actually clustering. The second, $-\frac{6 }{5} f_\nu\log \frac{1+z_\nu}{1+z}$, is due to reduced rate of growth of the dark matter perturbations in the presence of matter that doesn't cluster. Using 
\beq\label{eq:fnu}
    \Omega_\nu h^2=6 \times 10^{-4}\left(\frac{\sum m_\nu}{58 \,  \mathrm{meV}}\right) \to f_\nu \approx 4\times 10^{-3} \ ,
\eeq
the suppression of the matter power spectrum at $z=1$ is expected to be 
\beq
    P^{(\sum m_\nu = 58 \, {\rm meV})} (k\gg k_{\rm fs}, z ) \approx (1- 0.02) P^{(\sum m_\nu =0)}(k\gg k_{\rm fs}, z ) \ .
\eeq
Therefore, the signal we are looking for is a 2\% suppression of power on small scales around $z={\cal O}(1)$.

Galaxy surveys like DESI do not directly measure $P(k)$ and instead primarily measure the clustering of galaxies. The power spectrum of galaxy overdensity has an overall amplitude that depends on the details of galaxy formation, and the baryonic physics inherent in galaxy formation is understood with insufficient precision to directly extract the amplitude of $P(k)$ from these measurements. The best current measurements of the matter power spectrum come from gravitational lensing of the CMB. The CMB lensing convergence power spectrum $C_\ell^{\kappa \kappa}$  is given in the Limber approximation  by~\cite{Lewis:2006fu}
\beq
    C_\ell^{\kappa \kappa}\approx 2\pi^2 \ell \int_{\eta_*}^{\eta_0} \eta \mathrm{d} \eta \mathcal{P}_{\Psi}\left(\ell / (\eta_0-\eta) ; \eta\right)\left(\frac{\eta_*-\eta}{(\eta_0-\eta_*) (\eta_0-\eta)}\right)^2 \, ,
\eeq
where $\eta$ is the conformal time with $\eta_\star$ and $\eta_0$ denoting the times of recombination and $z=0$ respectively. We also defined $\mathcal{P}_{\Psi}$ as power spectrum of the Weyl potential, $\Psi$, which can be written in terms of the matter power spectrum as 
\beq
    \mathcal{P}_{\Psi}\left(k; \eta \right) =\frac{9 \Omega_m^2(\eta) H^4(\eta)}{8 \pi^2} \frac{P(k ; \eta)}{k} \, .
\eeq
Using the fact that the matter power spectrum is proportional to the primordial scalar amplitude $A_s$, we see that the amplitude of the CMB lensing power spectrum scales as 
\beq
    C_\ell^{\kappa\kappa} \propto (\Omega_m h^2)^2 A_s \left(1-0.02 \frac{f_\nu}{4\times 10^{-3}} \right) \ .
\eeq
Therefore, in order to measure a three-percent suppression of the lensing power spectrum, we must determine the physical matter density $\Omega_m h^2$ (where $h=H_0/ (100 ~\mathrm{km}\,\mathrm{s}^{-1}\,\mathrm{Mpc}^{-1})$ is the dimensionless Hubble constant) and the primordial scalar amplitude $A_s$ to much better than three-percent accuracy. 

%%%%%%%%%%%%%%%%%
\begin{figure}[t!]
    \centering
    \includegraphics[width=0.85\textwidth]{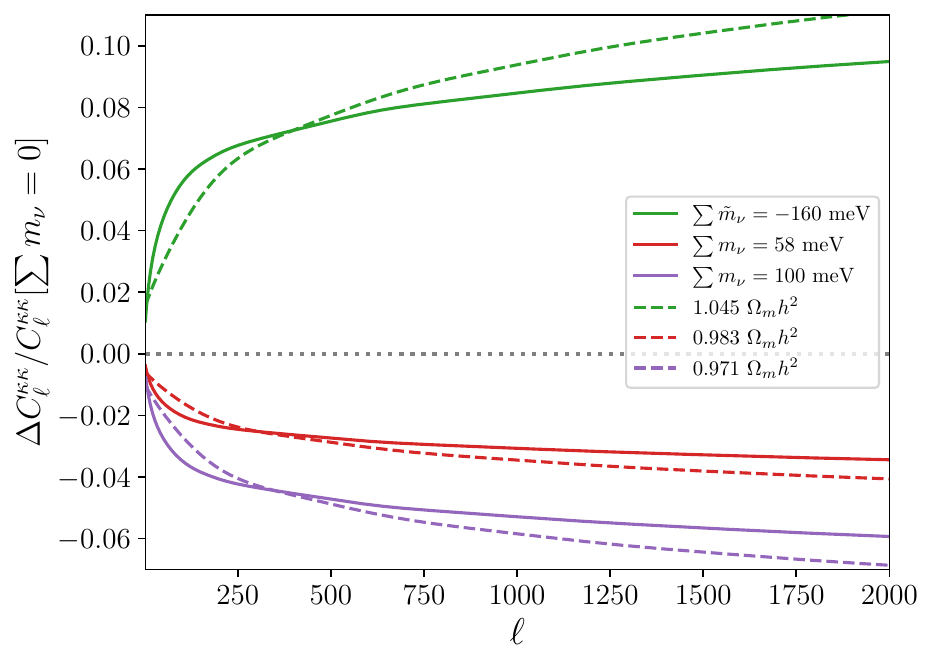}
    \caption{Comparison of the fractional change to the CMB lensing power spectrum from changes to $\Omega_m h^2$ and the introduction of a non-zero neutrino mass.}
    \label{fig:omegam}
\end{figure}
%%%%%%%%%%%%%%%%%%

The main impact of DESI on the cosmological neutrino mass constraint is to provide a precise measurement of $\omega_m \equiv \Omega_m h^2$ through the constraint on the expansion history from BAO. The impact of changing $\omega_m$ on the CMB lensing power spectrum is shown in Figure~\ref{fig:omegam} (for $\sum m_\nu = 0$ and compared to the change from introducing $\sum m_\nu > 0$). The reduction of $\omega_m$ by $1.7\%$ is roughly equivalent to introducing $\sum m_\nu = 58$ meV, which implies that a $2\sigma$ measurement of the minimum sum requires roughly 0.8\% precision in the measurement of $\omega_m$.

%%%%%%%%%%%%%%%%%%%%%%%%%%%%%%%%
\subsection{Negative Neutrino Mass}
\label{sec:Negative_mnu}
%%%%%%%%%%%%%%%%%%%%%%%%%%%%%%%

The physical sum of neutrino masses is of course restricted to be positive.  However, the combination of cosmological observables that we use to infer the mass of neutrinos are not restricted in this manner.  We show in this subsection that the CMB+DESI data in fact prefer a negative neutrino mass (already hinted at in eBOSS~\cite{eBOSS:2020yzd}), corresponding to increased matter clustering compared to a model with only massless neutrinos.

In order to measure the preference of cosmological data for negative neutrino mass, we require an implementation of the effects of neutrino mass that is allowed to take either sign.  The Boltzmann codes \texttt{CAMB}~\cite{Lewis:1999bs,Howlett:2012mh} and \texttt{CLASS}~\cite{Blas:2011rf} model neutrino mass in a way that is subject to the physicality constraint $\sum m_\nu > 0$.  We modified \texttt{CAMB} to include a new parameter, $\sum \tilde{m}_\nu$, which is designed to mimic the effects of neutrino mass, but which is not restricted to be positive.  Our new parameter simply scales the CMB lensing power spectrum in the same manner that would be expected from $\sum m_\nu$.  Specifically, we determine the fractional change $A_\ell(\sum \tilde{m}_\nu) \equiv C_\ell^{\kappa\kappa}[\sum m_\nu]/C_\ell^{\kappa\kappa}[\sum m_\nu=0]$ at fixed values of $H_0$, $\omega_m$, and $\omega_b$.  Once calibrated on positive values of neutrino mass, the effects of $\sum \tilde{m}_\nu$ can then be straightforwardly calculated for negative values as well.  In the $\Lambda$CDM+$\sum \tilde{m}_\nu$ cosmology, observables are computed with the physical $\sum m_\nu = 0$ and the CMB lensing power spectrum is computed as $C_\ell^{\kappa\kappa} = A_\ell(\sum \tilde{m}_\nu) C_\ell^{\kappa\kappa}[\sum m_\nu=0]$.  The temperature and polarization CMB power spectra are lensed using this modified CMB lensing power spectrum such that the set {$C_\ell^{TT}$, $C_\ell^{TE}$, $C_\ell^{EE}$, $C_\ell^{\kappa\kappa}$} is calculated self-consistently for each point in parameter space.

This prescription is very similar, though not identical, to the effects of the physical neutrino mass in the regime $\sum m_\nu > 0$.  In particular, the physical neutrino mass in the $\Lambda$CDM+$\sum m_\nu$ cosmology contributes to the non-relativistic matter density today $\Omega_m = \Omega_b + \Omega_c + \Omega_\nu$.  In our $\Lambda$CDM+$\sum \tilde{m}_\nu$ cosmology, there is no neutrino contribution to $\Omega_m$.  As a result, we anticipate that $\sum \tilde{m}_\nu$ should exhibit slightly weaker constraints than the physical $\sum m_\nu$ when measured using the same data combination.  To check this, we derive constraints on three cosmological models: a model with a physical neutrino mass $\Lambda$CDM+$\sum m_\nu$, a model with our parametrized neutrino mass restricted to positive values $\Lambda$CDM+$(\sum \tilde{m}_\nu > 0)$, and finally a model with our parametrized neutrino mass with no restriction on sign $\Lambda$CDM+$\sum \tilde{m}_\nu$.  We analyze each model using the same data combination.

Boltzmann calculations were carried out using our modified version of \texttt{CAMB}~\cite{Lewis:1999bs,Howlett:2012mh}.  We utilized the likelihood for CMB temperature and polarization from Planck's 2018 data release~\cite{Aghanim:2019ame}, along with the combination of ACT DR6~\cite{ACT:2023dou,ACT:2023kun} and Planck CMB lensing~\cite{Carron:2022eyg}, and DESI BAO~\cite{DESI:2024lzq,DESI:2024mwx,DESI:2024uvr}.  This combination of data is the same as that used by the DESI team to derive cosmological constraints~\cite{DESI:2024mwx}.  Our analysis was performed with \texttt{cobaya}~\cite{Torrado:2020dgo}, using the Markov chain Monte Carlo sampler adapted from CosmoMC~\cite{Lewis:2002ah,Lewis:2013hha} using the fast-dragging procedure~\cite{Neal:2005}.  Analyses were run until the Gelman-Rubin statistic was $R-1<0.01$. 

\begin{figure}[ht!]
    \centering
    \includegraphics[width=\textwidth]{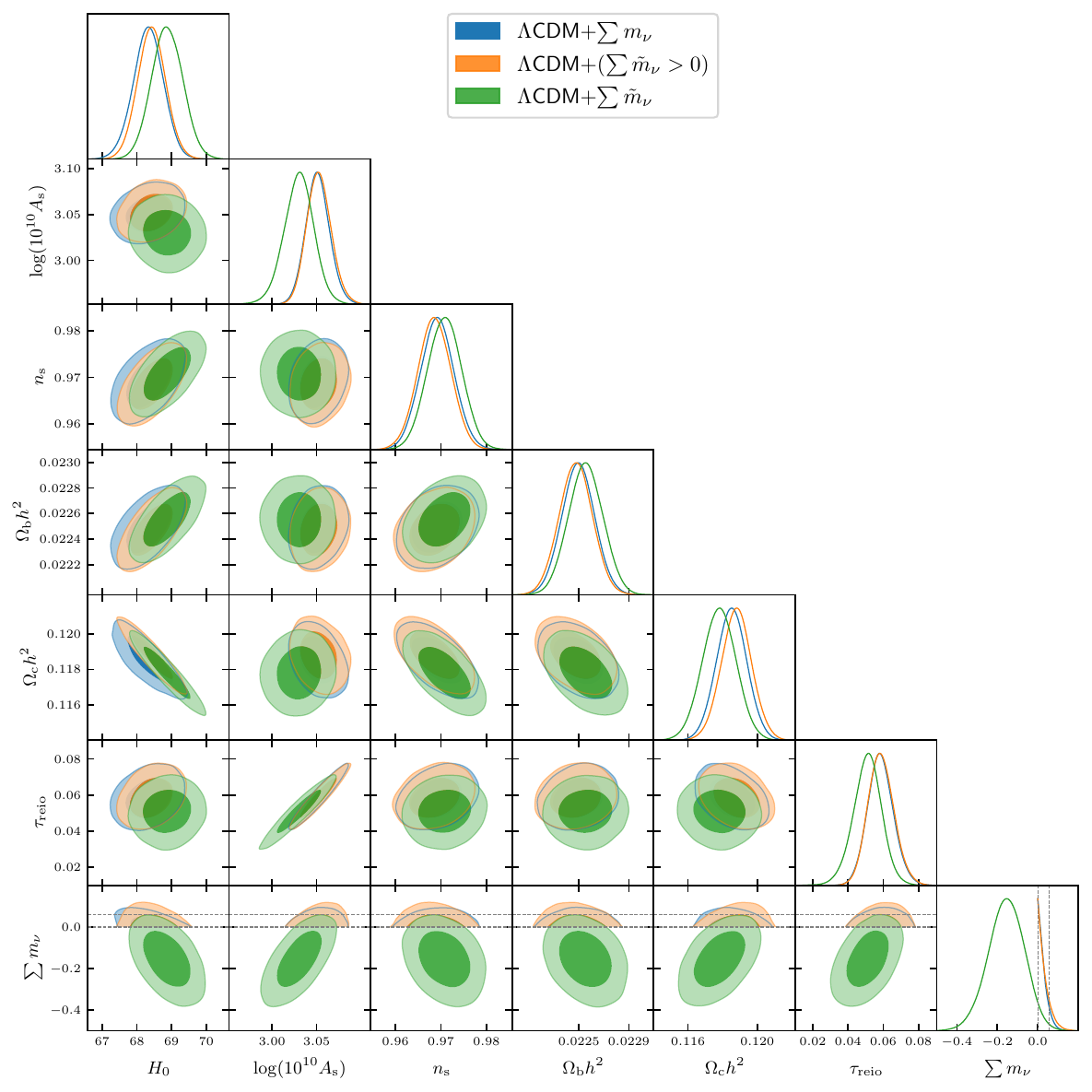}
    \caption{Triangle plot showing parameter constraints from Planck + ACT lensing + DESI BAO in three models described in the text and shown in Table~\ref{tab:constraints}.  For the purposes of this plot, we treat the physical neutrino mass and our parametrized version on the same footing.  Dashed lines show vanishing neutrino mass $\sum m_\nu = 0$ and the minimal sum of neutrino mass $\sum m_\nu = 58$~meV. Values of neutrino mass are reported in eV and $H_0$ in $\mathrm{km}\, \mathrm{s}^{-1}\, \mathrm{Mpc}^{-1}$.}
    \label{fig:triangle}
\end{figure}

The results are presented in Table~\ref{tab:constraints} and Figure~\ref{fig:triangle}. Notice that the parameter constraints in the $\Lambda$CDM+$\sum m_\nu$ and $\Lambda$CDM+$(\sum \tilde{m}_\nu > 0)$ models are nearly identical, showing only slightly weaker constraints on $\sum \tilde{m}_\nu$ as compared to the physical $\sum m_\nu$.  This excellent agreement justifies our prescription for modeling the effects of neutrino mass, with the slightly weaker constraints on $\sum \tilde{m}_\nu$ expected from the differing treatment of $\Omega_m$ in the two models.  Notice that in the $\Lambda$CDM+$\sum \tilde{m}_\nu$ model, the best-fit value for $\sum \tilde{m}_\nu$ is $-160$~meV, showing a preference for negative neutrino mass, and disfavoring even the minimal sum of neutrino masses inferred from flavor oscillation experiments at 3$\sigma$.  

We also note in passing that in the $\Lambda$CDM+$\sum \tilde{m}_\nu$ model, the best-fit value for $S_8\equiv \sigma_8(\Omega_m/0.3)^{0.5}$ is lower than in $\Lambda$CDM+$\sum m_\nu$ by about $1.5\sigma$ and has $40\%$ larger error bars (and is also smaller than the value inferred with Planck in the $\Lambda$CDM model, for which $S_8=0.830\pm0.013$~\cite{Planck:2018vyg}), representing a somewhat smaller $S_8$ tension~\cite{Abdalla:2022yfr} when neutrino mass is allowed to be negative.

%%%%%%%%%%%%%%%%%%%%%%
\begin{table}[t!]
    \begin{center}

\begin{tabular} { l  d e f}
\noalign{\vskip 3pt}\hline\noalign{\vskip 1.5pt}\hline\noalign{\vskip 5pt}
    & {\bf $\Lambda$CDM+$\sum m_\nu$} &  {\bf $\Lambda$CDM+$(\sum \tilde{m}_\nu>0)$} & {\bf $\Lambda$CDM+$\sum \tilde{m}_\nu$}\\
\noalign{\vskip 3pt}\cline{2-4}\noalign{\vskip 3pt}

 Parameter &  68\% limits &  68\% limits &  68\% limits\\
\hline
{\boldmath$\log(10^{10} A_\mathrm{s})$} & $3.051\pm 0.014            $ & $3.053^{+0.013}_{-0.014}   $ & $3.030\pm 0.017            $\\

{\boldmath$n_\mathrm{s}   $} & $0.9692\pm 0.0037          $ & $0.9686\pm 0.0036          $ & $0.9708\pm 0.0038          $\\

{\boldmath$100\theta_\mathrm{MC}$} & $1.04112\pm 0.00029        $ & $1.04111\pm 0.00029        $ & $1.04118\pm 0.00029        $\\

{\boldmath$\Omega_\mathrm{b} h^2$} & $0.02249\pm 0.00013        $ & $0.02248\pm 0.00013        $ & $0.02255\pm 0.00014        $\\

{\boldmath$\Omega_\mathrm{c} h^2$} & $0.11852\pm 0.00088        $ & $0.11880\pm 0.00088        $ & $0.11780\pm 0.00097        $\\

{\boldmath$\sum m_\nu \, (\sum \tilde{m}_\nu)            $} & $< 0.0741~(95\%~\mathrm{CL})                  $ & $< 0.0926~(95\%~\mathrm{CL})                  $ & $-0.156^{+0.093}_{-0.085}  $\\

{\boldmath$\tau_\mathrm{reio}$} & $0.0585\pm 0.0074          $ & $0.0588^{+0.0069}_{-0.0077}$ & $0.0510\pm 0.0083          $\\

\hline

$H_0                       $ & $68.33\pm 0.43             $ & $68.43\pm 0.40             $ & $68.87\pm 0.45             $\\

$\Omega_\mathrm{m} h^2     $ & $0.14131\pm 0.00084        $ & $0.14127\pm 0.00083        $ & $0.14036\pm 0.00092        $\\

$\sigma_8                  $ & $0.8175^{+0.0076}_{-0.0058}$ & $0.8246^{+0.0053}_{-0.0060}$ & $0.8123\pm 0.0078          $\\

$S_8\equiv \sigma_8(\Omega_m/0.3)^{0.5}$                      & $0.8212\pm 0.0096          $ & $0.827\pm 0.010            $ & $0.807\pm 0.013            $\\
\hline
\end{tabular}
\end{center}
\caption{Parameter constraints from Planck + ACT lensing + DESI BAO in the three models described in the text.  All constraints are given as $68\%$ limits, except for the upper limits on the neutrino mass when it is restricted to be positive, which are reported as $95\%$~CL.  Values of neutrino mass are reported in eV and $H_0$ in $\mathrm{km}\, \mathrm{s}^{-1}\, \mathrm{Mpc}^{-1}$.  In the $\Lambda$CDM+$\sum \tilde{m}_\nu$ model, the data favors a negative neutrino mass and disfavors the minimal physical neutrino mass of 58~meV at $3\sigma$.
}
\label{tab:constraints}
\end{table}
%%%%%%%%%%%%%%%%%%%%%%%%%%%%%%%%%

%%%%%%%%%%%%

\subsection{Influence of Cosmic Parameters}
\subsubsection*{Optical Depth}

The measurement of $A_s$ is limited by our understanding of the optical depth to reionization, $\tau$. Thomson scattering of CMB photons into and out of the line of site by free electrons present after reionization suppresses the amplitude of CMB fluctuations.  The observed amplitude of the CMB power spectrum is thereby reduced on small angular scales.  CMB observations primarily constrain the combination~\cite{Planck:2018vyg}
\beq
A_s e^{-2 \tau} = (1.884\pm 0.011)\times 10^{-9}  \ .
\eeq
This should be contrasted with the much less precise measurement of the primordial amplitude~\cite{Planck:2018vyg}
\beq
A_s  = (2.100 \pm 0.030) \times 10^{-9} \ .
\eeq
Noting that for these same analyses,
\beq
\tau =0.0544\pm 0.0073 \, ,
\eeq
the error on $A_s$ can be directly attributed to the error in $\tau$ and not the error in the measurement of $A_s e^{-2\tau}$.

\begin{figure}
	\centering
	\includegraphics[width=0.85\textwidth]{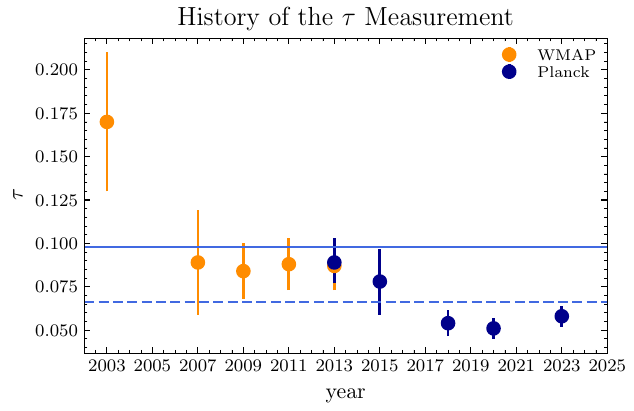}
	\caption{The historical measurement of the optical depth, $\tau$, from WMAP data~\cite{WMAP:2003ivt,WMAP:2006bqn,WMAP:2008lyn,WMAP:2010qai,WMAP:2012nax} and Planck~\cite{Planck:2013pxb,Planck:2015fie,Planck:2018vyg,Planck:2020olo,Tristram:2023haj} by year of publication. The horizontal solid (dashed) blue line indicates the central value of $\tau$ that would be to move the peak of the DESI+CMB the $\sum m_\nu$ posterior from -160 meV (0 meV) to 58 meV. 
    }
	\label{fig:tau}
\end{figure}

Of all the cosmological parameters defining $\Lambda$CDM, the optical depth is the most challenging to measure. For $\ell > 30$, its effects on the CMB are completely degenerate with $A_s$. It is only on large angular scales that the optical depth leaves a unique imprint, through the production of CMB polarization and the associated `reionization bump' in the polarization power spectrum. The history of these measurements, shown in Figure~\ref{fig:tau} has involved significant changes in the central value with relatively small changes in sensitivity. 

It is natural to wonder if the apparent measurement of $\sum m_\nu = -160$ meV could also be attributed to an error in the measurement of $\tau$. For this to be possible, we would need the true value of $A_s$ to be roughly 8.8\% larger, so that the current measurement of the lensing includes the expected suppression of $P(k)$ relative to $A_s$. This would require a value of the optical depth larger than that inferred from Planck $\tau = \tau_{\rm Planck}+ \delta \tau$, such that $2 \delta \tau \approx 0.088$. Using $\tau_{\rm Planck18} = 0.054$ and $\sigma_{\rm Planck18} =0.0073$, this would require 
\beq
    \tau_{\rm true} \geq 0.098  = \tau_{\rm Planck18}+ 6.0 \sigma_{\rm Planck18} \ \ .
\eeq
Similarily, if we take $\tau = 0.051 \pm 0.006$ or $\tau = 0.058 \pm 0.006$ from~\cite{Planck:2020olo} and~\cite{Tristram:2023haj}, we would require shifts of $7.8\sigma$ or $6.7\sigma$ respectively. For comparison, to shift $\sum m_\nu = 0$ to 58 meV only requires $A_s$ to be 2.5\% larger, which can be accomplished by a $\tau=0.066$ which is a 1.7$\sigma$ upward shift. Both lines are shown in Figure~\ref{fig:tau} and are consistent with some historical measurements; thus a systematic offset in the more recently inferred values of the optical depth is a plausible explanation for preference for negative neutrino mass. Yet, due to the magnitude of the difference it is unlikely to be the result of a statistical fluctuation.  

One of the key challenges with the optical depth is that it is very difficult to measure with ground-based surveys (although it is currently being pursued, for example, by the Cosmology Large Angular Scale Surveyor (CLASS) collaboration~\cite{Essinger-Hileman:2014pja,Eimer:2023esh}). The results of DESI alone point to the need for a confirmation of the Planck measurement of the optical depth, and in principle an improvement to the cosmic variance limit of $\sigma(\tau)= 0.002$. This would be possible with another satellite, such as LiteBird~\cite{LiteBIRD:2022cnt}. However, there is the more immediate potential of balloon-based observations which could reach similar levels of sensitivity~\cite{Errard:2022fcm}. Other longer term possibilities include using measurements of cross-correlations between the CMB and galaxy surveys to eliminate the need for an optical depth measurement~\cite{Yu:2018tem,Brinckmann:2018owf} or to use measurement the patchy kinetic Sunyaev-Zeldovich effect to constrain the physical model of reionization~\cite{Smith:2016lnt,Ferraro:2018izc,Alvarez:2020gvl}, both of which might be possible with CMB-S4~\cite{Abazajian:2019eic}.

\subsubsection*{Matter Content}

The measurement of the matter density $\omega_m$ is equally important to the measurement of $\sum m_\nu$ as the optical depth. The primary CMB directly determines $\omega_m$ through its influence on the height and locations of the acoustic peaks. This is, in part, why the CMB alone is capable of producing very stringent bounds on $\sum m_\nu$, e.g.~$\sum m_\nu < 240$ meV (95\%) from Planck TTTEEE + lensing~\cite{Planck:2018vyg}.

\begin{figure}
	\centering
	\includegraphics[width=0.85\textwidth]{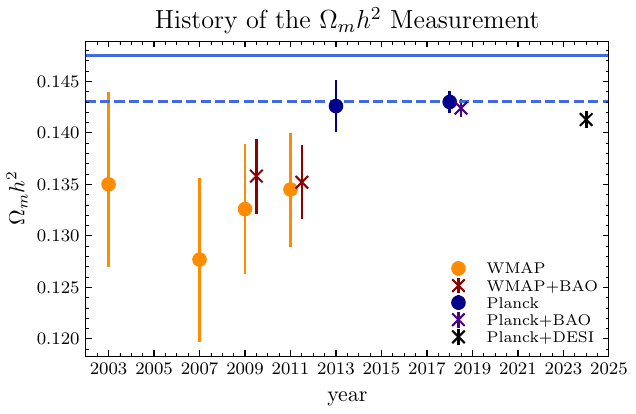}
	\caption{The historical measurement of the matter density, $\Omega_m h^2$, from WMAP data~\cite{WMAP:2003ivt,WMAP:2006bqn,WMAP:2008lyn,WMAP:2010qai} and Planck~\cite{Planck:2013pxb,Planck:2018vyg} by year of publication. The black Planck+DESI point is the result of our reanalysis of $\Lambda$CDM+$\sum m_\nu$ using the same priors as~\cite{DESI:2024mwx}. The horizontal solid (dashed) blue line indicates the central value of $\Omega_m h^2$ that would be required to move the peak of the CMB+DESI $\sum m_\nu$ posterior from -160 meV (0 meV) to 58 meV. }
	\label{fig:omegam_hist}
\end{figure}

Improvements in the measurement to $\omega_m$ beyond the CMB has been driven by BAO measurements, most recently with DESI. As shown in Figure~\ref{fig:omegam_hist}, the BAO has played a significant role in reducing uncertainty, but has been consistent with the measurements from the CMB data on which the BAO is calibrated. Like the measurement of the optical depth, there was a significant improvement from WMAP to Planck. However, unlike $\tau$, the Planck measurements of $\omega_m$ have been stable with the inclusion of more data, including from polarization and the BAO.

The measurement of $\omega_m$ needs to be accurate to less than 0.8\% in order to permit a reliable measurement of $\sum m_\nu$. While this is a high standard, we have the benefit that $\omega_m$ will be measured using a number of different CMB surveys that can be combined with several large-scale structure (LSS) surveys. Any large shifts in $\omega_m$ due to systematic effects should be different for different surveys and thus from planned measurements alone, we should be able to determine a robust value of $\omega_m$ and/or identify systematic issues. This is in sharp contrast to the optical depth, of which Planck is currently the only measurement at the needed accuracy, and it is unclear if near term observations will reproduce or exceed their sensitivity.

It is well known that introducing dynamical dark energy, e.g.~in the form of $w_0 \neq -1$ and $w_a \neq 0$, significantly weakens~\footnote{Imposing $\sum m_\nu > 0$, it has been observed that constraints from current data on neutrino mass can tighten when marginalizing over some models of non-phantom dynamical dark energy~\cite{Vagnozzi:2018jhn,Berghaus:2024kra}. While the Cramer-Rao bound requires that the statistical uncertainty must increase, a shift of the central value to more negative values could explain this behavior.} the neutrino mass constraints~\cite{Allison:2015qca}. This is for the simple reason that if we allow for more free parameters in the expression for $H(z)$ at low redshifts, we cannot measure $\omega_m$ at the accuracy needed to determine $\sum m_\nu$. However, this will typically require fairly significant changes to the content and history of the universe. Leaving the content of the universe fixed, we will see that the neutrino mass signal can be explained with changes to the micro-physics in the neutrino and/or dark sector that otherwise leave the rest of cosmological history intact.

\subsubsection*{CMB Lensing}

Weak gravitational lensing of the CMB perturbs the path of photons, so that the apparently location on the sky is perturbed from the true direction $\hat{n}^{\prime}=\hat{n}+\vec{\alpha}(\hat{n})$, where $\vec\alpha(\hat n)$ is deflection angle~\cite{Lewis:2006fu}. Since the gravitational lensing is time-independent on the scales of observations, the maps of the CMB temperature anistropies (for example) are also modified by the same effect,
\beq
    T_{\text {lensed }}(\hat{n})=T_{\text {unlensed }}(\hat{n}+\vec{\alpha}(\hat{n})) \ .
\eeq
The deflection angle is related to the gravitational potential via the lensing potential $\phi(\hat n)$, via $\vec \alpha = \nabla_{\hat n} \phi$ and
\beq
    \phi(\hat{n}) \equiv-2 \int_{\eta_{\star}}^{\eta_0} \mathrm{d} \eta \frac{\eta-\eta_{\star}}{(\eta_0-\eta_{\star})(\eta_0-\eta)} \Psi\left((\eta_0-\eta) \hat{\mathbf{n}}, \eta\right) \ ,
\eeq
where $\Psi$ is the Weyl potential and $\eta_\star$ is the conformal time of CMB last scattering. 

We can understand the main influence of lensing on the CMB by Taylor expanding
\beq
    T_{\text {lensed }}(\hat{n}) \approx T_{\text {unlensed }}(\hat{n}) + \nabla_{\hat n} T \cdot \nabla_{\hat n} \phi + \mathcal{O}(\phi^2) \, .
\eeq
For a small patch of sky, we can Fourier transform $\hat n \to \vec \ell$ so that the dot product is replaced with a convolution
\beq
    T_{\text {lensed }}(\vec \ell) \approx T_{\text {unlensed }}(\vec \ell) - \int \frac{d^2\vec L}{2\pi} \, \vec L \cdot (\vec \ell -\vec L) T_{\text {unlensed }}(\vec \ell -\vec L ) \phi(\vec L) +  \mathcal{O}(\phi^2) \, .
\eeq
This will induce a non-vanishing correlation between different Fourier modes,
\begin{align}
    &\left\langle T_{\text {lensed }}(\vec \ell) T_{\text {lensed }}( \vec L - \vec \ell) \right\rangle_T =
    \delta(\vec L) C^{TT\mathrm{,unlensed}}_{\ell} \nonumber \\
    & \qquad + \frac{1}{2\pi} \left[ (\vec L - \vec \ell)\cdot \vec L C_{|\vec L-\ell|}^{TT\mathrm{,unlensed}}+\vec \ell \cdot \vec L C_{\ell}^{TT\mathrm{,unlensed}}\right] \phi(\vec L) + \mathcal{O}(\phi^2) \, ,
\end{align}
where $C_{\ell}^{TT\mathrm{,unlensed}}$ is in the unlensed temperature power spectrum, and the subscript $T$ on the left-hand side refers to an ensemble average over the unlensed CMB temperature realization. As the $\vec L \neq 0$ correlations would vanish without lensing, we can reconstruct $\phi(\vec L)$ from the presence of these correlations~\cite{Hu:2001kj}. Estimating the CMB lensing power spectrum can therefore be achieved by measuring the temperature four-point function. Lensing also induces a measurable smoothing effect on the acoustic peaks of the CMB power spectrum, from convolving the unlensed power spectrum with the lensing power spectrum at second order.

Once the lensing potential is reconstructed, it can be used to calculate the power spectrum of lensing, remove lensing from the CMB maps~\cite{Seljak:2003pn,Green:2016cjr,Hotinli:2021umk}, and/or cross-correlate with other data. For the neutrino mass, the only piece of information we need it the power spectrum of the lensing map $C_L^{\phi \phi}$. This is the same information that is contained in the connected trispectrum of the temperature, as $\phi(\vec L)$ was determined from a temperature two-point function. As shown in Figure~\ref{fig:omegam}, $\sum m_{\nu }= 58$ meV causes a roughly 2-3\% suppression of the lensing power, while $\sum m_\nu = - 160$ meV is a 6-9\% enhancement.

The reconstruction of the lensing map is a non-trivial process that could be influenced by other effects that correlate modes in the temperature maps. For example, it is known that the non-Gaussian statistics of unresolved foregrounds can induce biases in these maps~\cite{vanEngelen:2013rla}. Furthermore, these same correlations are relevant to the covariance of the primary CMB and thus are important for measurements of any other cosmological parameters. Yet, it is also noteworthy that the neutrino mass measurement not sensitive to non-linear effects in the matter power spectrum. Using current CMB data, the lensing map is too noisy to resolve modes that are strongly influenced by non-linear evolution. Yet, even with future data, such as from CMB-S4, these modes can be removed from the analysis with no loss of sensitivity to $\sum m_\nu$~\cite{Green:2021xzn}.

%%%%%%%%%%%%%%%%%%%%%%%%%%%%%%%%%%%%%%%
\section{Vanishing Neutrino Mass}\label{sec:zero}
%%%%%%%%%%%%%%%%%%%%%%%%%%%%%%%%%%%%%%%

In this section, we will explore mechanisms for eliminating the signal of $\sum m_\nu \geq 58$ meV, while being consistent with $\sum m_\nu \geq 0$. The common element of all these models is that we will reduce or eliminate the suppression of power by directly altering the behavior of the neutrinos. In the next section, we will consider changes to the growth of structure beyond just the neutrinos, which could allow for an apparent enhancement of structure, which might be interpreted as $\sum m_\nu <0$.

\subsection{Decays}

Perhaps the most obvious way to reconcile a cosmological indication of $\sum m_\nu = 0$ with the nonzero masses implied by neutrino oscillations is if massive neutrinos decay into massless degrees of freedom on cosmological timescales. While the two heaviest neutrino mass eigenstates are already unstable within the Standard Model, their lifetimes are far greater than the age of the universe ($\tau_\nu \propto \langle h \rangle^4/ m_\nu^5$, where $\langle h \rangle \simeq 246$~GeV is the vacuum expectation value of the Higgs field). Neutrino decays on cosmologically relevant timescales would therefore be unambiguous evidence of new physics, above and beyond the origin of neutrino masses. 

While decays involving photons are strongly constrained by CMB spectral distortions~\cite{Aalberts:2018obr}, decays into dark radiation (and either an active or sterile neutrino) are consistent with current limits over a wide range of lifetimes. A lower bound comes from the requirement that the decays and inverse decays of relativistic neutrinos do not prevent free streaming, $\tau_\nu \gtrsim 4 \times 10^6 \, {\rm s} \, (m_\nu / 0.05 \, {\rm eV})^5$~\cite{Barenboim:2020vrr}. On the upper end, the maximum neutrino lifetime that can erase the cosmological signal of neutrino masses depends on the mass spectrum~ \cite{Chacko:2019nej,Chacko:2020hmh,Escudero:2020ped,FrancoAbellan:2021hdb,Escudero:2022gez}. For the minimum masses implied by neutrino oscillations, the lifetime of the massive neutrinos should be roughly an order of magnitude shorter than the age of the universe, $\tau_\nu \lesssim 4 \times 10^{16} \, {\rm s}$. For the sum of neutrino masses to be observable at \mbox{KATRIN} (sensitive to $m_{\nu_e}$ as small as 0.2~eV~\cite{KATRIN:2021uub}, which translates to $\sum m_\nu \sim 0.6$~eV), the maximum lifetime of all the active neutrinos should be around two orders of magnitude smaller, $\tau_\nu \lesssim 4 \times 10^{14} \, {\rm s}$.

There are a variety of possible decay modes. Two-body decays of massive neutrinos necessarily proceed into a fermion and a boson, with the former either an active or sterile neutrino, and the latter a scalar $\phi$ or vector $Z'$. As the masses of the bosons increase, the two-body decay channels close and the bosons instead mediate three-body decays into active and sterile neutrinos. As the viable parameter space for three-body decays is considerably more constrained, here we will restrict our attention to the two-body decays.

In the neutrino mass basis, decays into a (pseudo)scalar arise via couplings of the form 
\beq \label{eq:scalarlag}
\mathcal{L}_\phi \supset \frac{\lambda_{i j}}{2} \bar{\nu}_i \nu_j \phi+ \frac{\tilde \lambda_{i j}}{2} \bar{\nu}_i \gamma_5 \nu_j \phi+{\rm h.c. }  \hspace{1cm} (i,j = 1,\dots 4) \ ,
\eeq
where $i = 1,2,3 (4)$ denote the primarily active (sterile) neutrino mass eigenstates; for definiteness we assume the neutrinos are Majorana. Assuming the lightest active or sterile neutrinos are much lighter than the heavy neutrinos, the corresponding lifetime for decay via the pseudoscalar coupling is \cite{Escudero:2020ped}
\beq
\tau(\nu_i \rightarrow \nu_j \phi) \simeq 7 \times 10^{17} \,{\rm s} \times \left(\frac{0.05 \, {\rm eV}}{m_{\nu_i}} \right) \left( \frac{10^{-15}}{\tilde \lambda_{ij}^2} \right)^2 \,.
\eeq
For two-body decays into active neutrinos to reconcile oscillation splittings with a cosmological measurement of $\sum m_\nu = 0$, necessarily $m_{\nu_3} \approx 0.05$ eV. Erasing the energy density in massive neutrinos without spoiling free streaming then implies $4 \times 10^{-15} \lesssim \lambda, \tilde \lambda \lesssim 4 \times 10^{-10}$~\cite{Farzan:2002wx,Chacko:2003dt,Friedland:2007vv,Archidiacono:2013dua,Baumann:2016wac}. The situation is analogous for decays into sterile neutrinos, although in this case the overall mass scale of active neutrinos may be significantly increased \cite{FrancoAbellan:2021hdb}. 

While the dimensionless couplings required to erase the cosmological neutrino mass signal are small, they are nicely compatible with expectations from UV-complete models. For example, models with spontaneously broken global horizontal lepton flavor symmetries \cite{Gelmini:1983ea} give rise to a goldstone mode coupling to neutrinos as in Eq.~(\ref{eq:scalarlag}). In such models the off-diagonal pseudoscalar couplings $\tilde \lambda_{ij}$ are generated via mixing between heavy sterile and light active neutrinos of order $\tilde \lambda_{ij} \simeq \sqrt{m_{\nu_i} m_{\nu_j}}/f$, where $f$ is the scale of spontaneous symmetry breaking. The desired size of $\tilde \lambda$ corresponds to $50 \, {\rm MeV} \lesssim f \lesssim 5$ TeV, implying new physics associated with neutrino mass generation around the TeV scale.

Alternately, decays into a vector arise via couplings of the form
\beq \label{eq:vectorlag}
\mathcal{L}_{Z'} \supset \frac{g_{ij}^L}{2} Z_\mu' \, \bar \nu_i \gamma^\mu P_L \nu_j   + \frac{g_{44}^R}{2} Z_\mu'  \, \bar \nu_4 \gamma^\mu P_R \nu_4 +  {\rm h.c.} \hspace{1cm} (i,j = 1,\dots 4) \ ,
\eeq
which set a lifetime via two-body decays of order 
\beq
\tau(\nu_i \rightarrow \nu_j Z') \simeq 7 \times 10^{17} \, {\rm s} \times \left(\frac{0.05 \, {\rm eV}}{m_{\nu_i}} \right)^3 \left( \frac{m_{Z'}/g^L_{ij} }{50 \, {\rm TeV}}\right)^2 \,.
\eeq
For two-body decays into active neutrinos to erase the cosmological neutrino mass signal without spoiling free streaming requires $100 \, {\rm MeV} \lesssim m_{Z'}/g^L \lesssim 10 \, {\rm TeV}$, along with $m_{Z'} \ll m_{\nu_i}$. The situation is analogous for decays into sterile neutrinos, modulo the greater freedom in the active neutrino masses.

As in the scalar case, the dimensionless couplings required to erase the cosmological neutrino mass signal are nicely compatible with expectations from UV-complete models. For instance, in a model with a gauged lepton flavor symmetry such as $U(1)_{L_\mu-L_\tau}$ broken at a scale $f$, we have $f = m_{Z'} / g^L$ and the preferred range of decay couplings once again suggests new physics around the TeV scale. The preferred range of couplings and masses is also compatible with current limits, with the most stringent direct bounds $m_{Z'} / g^L > 1.3$~GeV coming from monolepton + missing energy searches at the LHC \cite{Ekhterachian:2021rkx}. 

\subsection{Annihilation}

The cosmological neutrino mass signal may alternately be erased if the cosmological population of massive neutrinos annihilates away into light states at late times \cite{Beacom:2004yd}. For simplicity, consider the case of a single light (pseudo)scalar coupling to neutrinos, as in Eq.~(\ref{eq:scalarlag}). Whereas neutrino decays require off-diagonal couplings in the mass basis, annihilation is efficient even when the largest couplings are diagonal. For annihilations to effectively deplete the relic neutrino abundance, the couplings $\lambda, \tilde \lambda$ should be large enough to keep $\phi$ in thermal equilibrium with neutrinos until after the neutrinos become non-relativistic, at which point the neutrinos annihilate efficiently via $\nu \nu \rightarrow \phi \phi$. The relic neutrino population is effectively erased provided $\lambda, \tilde \lambda \gtrsim 10^{-5}.$ However, such large couplings bring $\phi$ into thermal equilibrium before big bang nucleosynthesis (BBN), and the model is ruled out by a combination of free-streaming requirements and CMB bounds on $N_{\rm eff}$.

However, mild variations on this scenario remain consistent with current cosmological bounds \cite{Farzan:2015pca}. One natural possibility is for the active neutrinos to coannihilate into sterile neutrinos via a scalar or pseudoscalar $\phi$ through the couplings in Eq.~(\ref{eq:scalarlag}). Avoiding efficient coannihilation while neutrinos are still in thermal equilibrium implies $m_\phi \lesssim {\rm MeV}$, while ending coannihilation before recombination implies $m_\phi \gtrsim {\rm eV}$. Within this mass range, efficient conversion requires $\lambda, \tilde \lambda \gtrsim 5 \times 10^{-11} \times \left( \frac{m_\phi}{\rm keV} \right)^{1/2}$, while preserving free streaming at recombination requires $\lambda, \tilde \lambda \lesssim 5 \times 10^{-3} \times \left( \frac{m_\phi}{\rm keV} \right)$.

The above bounds are based on the direct coupling of active neutrinos to $\phi$. These are significantly weakened if the active neutrino couples to $\phi$ via light right handed neutrinos. In this case, in the early universe, the mixing of relativistic active neutrinos to the right handed neutrino is suppressed by the small neutrino mass, suppressing annihilation at early times. At low redshift, the mixing of non-relativistic neutrinos is unsuppressed, leading to enhanced annihilation that can also explain this signal. We briefly comment on this possibility in Section  \ref{subsec:timevaryingmass}.

\subsection{Cooling and Heating}

The origin of the neutrino mass signal in the matter power spectrum is that the neutrinos are cold enough to redshift like matter, but not cold enough to cluster like matter. Naturally, we could eliminate this signal by either heating or cooling the neutrinos. However, any large change to the temperature would have to come after recombination, as the measurement $N_{\rm eff} = 2.99 \pm 0.33$ (95\%)~\cite{Planck:2018vyg} is in precise agreement with the neutrino density predicted by the Standard Model~\cite{Mangano:2005cc,EscuderoAbenza:2020cmq,Akita:2020szl,Froustey:2020mcq,Bennett:2020zkv,Bond:2024ivb} and inferred from BBN~\cite{Fields:2019pfx}.

Cooling the neutrinos can be an effective strategy if they can be cooled enough to reduce the free-streaming scale below the nonlinear scale, or equivalently $k_{\rm fs} > k_{\rm NL} = {\cal O}(1) \, h \, {\rm Mpc}^{-1}$. Recall that the free-streaming scale is defined by 
\beq
    k_{\rm fs} = \sqrt{\frac{3}{2}}\frac{aH}{c_\nu} \ ,
\eeq
where the neutrino speed in the Standard Model is given by 
\beq
    c_\nu = \frac{\langle p_\nu \rangle}{m_\nu} = \frac{3 T_\nu}{m_\nu} \approx 1.0 \times 10^{-2} \left(\frac{50 \, {\rm meV}}{ m_\nu} \right) \, (1+z)   \ ,
    \label{eq:NeutrinoSpeedSM}
\eeq
As a result, the free-streaming scale as a function of the neutrino temperature is 
\beq
    k_{\rm fs}(z) = 0.04 \, h \, {\rm Mpc}^{-1}  \, \left(\frac{\sum m_\nu}{58 \, {\rm meV}}\right) \times \left(\frac{1.95 \, {\rm K}}{T_\nu(z=0)}\frac{1}{1+z}\right)  \ .
    \label{eq:kfs_nu_cooling}
\eeq
The role of the $z$-dependence puts a somewhat non-trivial requirement on $T_\nu$. At a minimum, if we have $k_{\rm fs}(z \approx 100) > 0.1 \, h \, {\rm Mpc}^{-1}$, then we could expect the neutrinos to cluster on the scales in the linear regime of our late time observations. Less conservatively, we require $k_{\rm fs}(z=0) > 0.1 \, h \, {\rm Mpc}^{-1}$. Together, these imply we need to cool the neutrinos by a factor of 10 to 1000 at redshifts $z< 1000$ to avoid the neutrino mass signal.

Solving for the coupled linear evolution of the dark matter, baryon, and neutrinos numerically (see Appendix~\ref{app:A}), Figure~\ref{fig:nu_cooling} shows the suppression as a function of the neutrino temperature as $z=0$, $T_{\nu}$ for the minimum sum of neutrino masses, $\sum m_\nu = 58$ meV. From these numerical results, we can conclude that $T_\nu < {\cal O}(1) \times 10^{-2}$ K at $z=0$ is sufficient to move the free-streaming signal to the non-linear regime, assuming that neutrino cooling occurs near $z =  100$.

A natural mechanism for cooling the neutrinos is through interactions with dark matter. The dark matter is cold and therefore is a natural heat sink for the neutrinos. It is straightforward~\cite{Green:2021gdc} to couple a right-handed neutrino, $N$, to dark matter, $\chi$ at low-redshifts through a light mediator $\phi$, 
\beq
\mathcal{L} \supset g_N \phi N N+g_\chi \phi \chi \chi+m^2 \phi^2+m_N N N+\lambda h L N+m_\chi \chi \chi \ .
\label{Eqn:ToyModel}
\eeq 
Scattering between the dark matter and neutrinos scales as $T_\nu^{-6}$ and thus avoids the constraints at earlier times (and higher temperatures) from BBN and the CMB~\cite{Nollett:2014lwa}. 

In order to cool the neutrinos and reproduce the clustering in a $\sum m_\nu =0$ universe, it is important that the scattering between neutrinos and dark matter is ineslatic.  This could be achieved through a number of mechanisms such a additional dark radiation coupled to $\chi$ or having nearly degenerate states associated with $\chi$ (like would occur with atomic dark matter, for example). This allows the dark matter to absorb energy from the neutrinos and allows for $T_\nu$ to decrease. In the above model, $g_N \sim g_\chi \approx 10^{-7}$ is sufficient to bring these two sectors into equilibrium at $z \lesssim 100$~\cite{Green:2021gdc} and any efficient process for absorbing the neutrino's energy would lead to an effecive $\sum m_\nu = 0$ signal.

One could also consider the case where $\chi$ is a single particle sub-component of the dark matter with total energy fraction $f_\chi$. Without any additional light states, the scattering between $\nu$ and $\chi$ is purely elastic. In this scenario, the effect of the coupling is create a neutrino-dark matter fluid, much like the photon-baryon fluid that fills the universe before recombination, with a free-streaming scale:
\beq
    k_{\mathrm{fs}} \approx 0.05 \, h \, \mathrm{Mpc}^{-1} \times\left(\frac{f_\chi+f_\nu}{f_\nu}\right)^{1 / 2}\left(\frac{\sum m_\nu}{58~\mathrm{meV}}\right) .
\eeq
The amplitude of the suppression on scales $k \gg k_{\rm fs}$ is proportional to $f_\chi + f_\nu$, the total energy fraction in this fluid. As a result, even if we could couple to all the dark matter so that $k_{\mathrm{fs}} = 0.8 \, h \, \mathrm{Mpc}^{-1}$, the suppression is large enough to be constrained by the Lyman-$\alpha$ forest~\cite{SDSS:2004kjl,Viel:2005qj,Boyarsky:2008xj,Xu:2018efh} or counts of satellite galaxies~\cite{Nadler:2019zrb,Maamari:2020aqz}.

\begin{figure}
	\centering
	\includegraphics[width=0.85\textwidth]{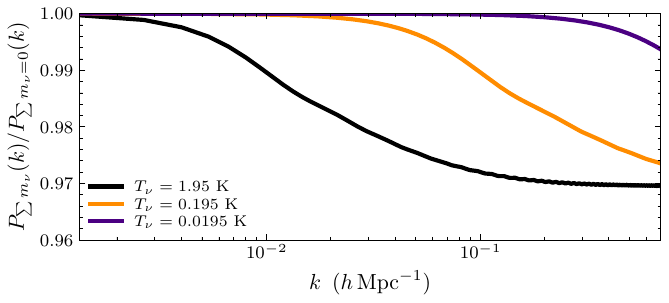}
	\caption{Suppression of $P_m(k)$ for $\sum m_\nu = 58$ meV and various neutrino temperatures at redshift zero, $T_\nu(z=0)$. As the suppression is a percent level effect, it will only be observable in the linear regime $k < 0.1 \, h \, {\rm Mpc}^{-1}$. We see that cooling to $T_\nu < 0.02$ K, or cooling by factor of 100, is sufficient to eliminate the signal of free-streaming neutrinos.}
	\label{fig:nu_cooling}
\end{figure}

Heating the neutrinos to avoid the suppression of matter clustering requires that the neutrino speed, shown in Eq.~\eqref{eq:NeutrinoSpeedSM}, remain near unity throughout cosmic history.  This could be achieved by increasing $T_\nu$ by a factor of $\sim 100$ in the regime $1000 \gtrsim z \gtrsim 100$; however, this would correspond to increasing the energy density of the cosmic neutrino background (C$\nu$B) by at least the same factor (assuming no change to the number density of neutrinos).  The extra energy density acquired by neutrinos needs to be transferred from another component, with the dark matter serving as the natural candidate during the matter-dominated era.  A transfer of energy from the dark matter to the C$\nu$B will have similar cosmological effects as models of dark matter decaying into dark radiation, which are subject to constraints from observations of the matter power spectrum and of the CMB that arise from a larger late-time integrated Sachs-Wolfe effect as compared to a standard cosmological history~\cite{Kofman:1986am,DeLopeAmigo:2009dc,Audren:2014bca,Poulin:2016nat}.  Current constraints set an upper limit of about 4\% of dark matter decaying into radiation after recombination~\cite{Poulin:2016nat}, comparable to the fraction of energy density that would need to be transferred from dark matter to heat the C$\nu$B in order to keep neutrinos relativistic until the present time.

\subsection{Time Varying Mass}

\label{subsec:timevaryingmass}

The tension between the DESI data and the laboratory measurement of neutrino masses can also be alleviated if the mass of the neutrino is not a constant in either time or space. For example, it might be the case that neutrinos had a smaller mass in the early universe (until around $z \sim 10$) but then subsequently had their mass change by $\mathcal{O}\left(1\right)$, as suggested in \cite{Fardon:2003eh, Lorenz:2018fzb, Lorenz:2021alz}. Alternately, it could be the case that the neutrino is a chameleon which acquires a larger mass near high density matter~\cite{Davoudiasl:2018hjw} but is otherwise lighter in the low density of the cosmos that is relevant to DESI and CMB lensing. For the purposes of illustration of this concept, in this paper, we study the possibility that the neutrino mass evolved in time and leave further exploration of potential chameleonic nature of neutrinos for future work.  

\begin{comment}
{\bf The text around this equation needs to be fleshed out}
The DESI data can also be explained if the neutrinos did not have a mass in the early universe (till around $z \sim 10$) but 

\beq
    P_{\sum m_\nu} (k\gg k_{\rm fs}, z ) \approx  \left(1 - 2 f_\nu  -\frac{6 }{5} f_\nu\log \frac{1+z_\nu}{1+z} \right) P_{\sum m_\nu =0}(k\gg k_{\rm fs}, z )
\eeq
\end{comment}

To realize the phenomenology of lower mass neutrinos that become more massive around $z \sim 10$, consider the following terms of the Lagrangian \eqref{Eqn:ToyModel}: 
\begin{equation}
\mathcal{L} \supset y h L N + g_N \phi N N + m^2 \phi^2 \ .
\end{equation}
We take the Yukawa coupling $y \sim \frac{\text{10 meV}}{\langle h \rangle}$ so that the neutrino's Dirac mass is comparable to the current neutrino mass $\sim 10$ meV (per neutrino). Observe that when $g_N \phi \gg y \langle h\rangle$ the phenomenology is identical to that of the conventional ``see-saw'' mechanism and thus the neutrino mass will be light. When $g_N \phi \ll y \langle h\rangle$, the Dirac mass will dominate and equal the desired present day value. The  cosmological evolution of $\langle \phi \rangle$ naturally leads to such a change due to the fact that $\langle \phi \rangle$ is sourced by the C$\nu$B, whose number density drops as the universe expands. 

To illustrate this dynamic, let us pick some example numbers. Suppose we assume that the neutrino mass was around $\sim 1$ meV in the early universe. Such neutrinos would be relativistic until $z \sim 10$. When they are relativistic, the C$\nu$B sources  $\langle\phi\rangle \sim g_N \frac{\text{meV}^3}{m^2}$ \cite{Green:2021gdc}, independent of the temperature of the neutrinos. Once the neutrinos become non-relativistic, $\langle \phi \rangle$ scales with the number density of the  C$\nu$B and we get $\langle \phi \rangle \sim  g_N \frac{T^3}{m^2}$. When $\langle \phi \rangle$ drops, the neutrinos become more massive, approaching their Dirac mass. The main constraint on this scenario is the bound $g_N \lesssim 5 \times 10^{-8}$ in order to ensure that the neutrinos do not annihilate into $\phi$ when they are light (in fact, if they do, the situation reduces to the annihilation scenarios considered earlier). Setting $g_N \sim 10^{-8}$ and $m \sim 10^{-12}$ eV, we see that at early times the neutrino mass is around  $\sim 1$ meV. These neutrinos become non-relativistic around $z \sim 10$. The subsequent drop in $\langle \phi \rangle$ raises the neutrino mass to around $\sim 20$ meV today (per neutrino).

\subsection{Mirror Sectors and Relation to the Hubble Tension}

The deviation from $\Lambda$CDM for $\sum m_\nu$ is roughly consistent with the suggestion that new physics might only impact dimensionful parameters~\cite{Cyr-Racine:2021oal,Ge:2022qws}. The CMB and LSS directly measure dimensionless quantities (angles, redshifts) and thus are not directly related to dimensionful quantities like $H_0$ and $\sum m_\nu$. This idea was put forward in Refs.~\cite{Cyr-Racine:2021oal,Ge:2022qws} to explain the Hubble tension. They realized this concept by introducing a mirror of the Standard Model in the dark sector, such that the gravitational signals remained unchanged but the Standard Model densities could be rescaled.

Naturally, such a model could also easily explain the apparent $\sum m_\nu \approx 0$, by having massless neutrinos in the hidden sector. This would leave the other gravitational signals unchanged, but reduce the total gravitational influence of the Standard Model neutrinos. This dilutes $\sum m_\nu$ by the fraction of matter in the hidden sector to the mirror sector, and thus requires the Standard Model to be a small component of the total matter density. Unlike some of the other solutions to $\sum m_\nu$, this requires an order one change to the universe and thus is difficult to make compatible with all observations. For example, BBN is sensitive to the physical baryon density and thus is not compatible with the simplest implementations of this idea.

Interestingly, the suggestion that there could be multiple copies of the Standard Model with different mass parameters is a natural consequence of several recent mechanisms for solving the hierarchy problem~\cite{Arkani-Hamed:2016rle,Chacko:2016hvu,Chacko:2018vss}. However, these hidden sector typically increase $\Neff > 3.044$ and $\sum m_\nu > 58$ meV. Without fine tuning these models to take the form of those described in Refs.~\cite{Cyr-Racine:2021oal,Ge:2022qws}, observations that favor $\sum m_\nu < 58$ meV severely constrain these models.

%%%%%%%%%%%%%%%%%%%%%%%%%%%%%%%%%%%%%%%
\section{Negative ``Neutrino Mass"}\label{sec:negative}
%%%%%%%%%%%%%%%%%%%%%%%%%%%%%%%%%%%%%%%

The possibility of an apparent measurement with $\sum m_\nu < 0$ would be most naturally explained by an increase in the amount of clustering in the late universe, or at least an apparent increase as measured through gravitational lensing of the CMB. Even if neutrinos were truly massless $\sum m_\nu = 0$, this would require a change to the formation of structure or the statistical properties of the CMB. Such a mechanism could also erase the signal from conventional massive neutrinos and thus need not involve a change to the neutrino sector at all. In this section, we will explore representative examples of how this signal could arise. We will consider physically increasing the amount of clustering through a new long range force, and creating a apparent increase in lensing through changes to the statistics of the primordial density fluctuations. Both classes of ideas lead to observable consequences that may already be testable with existing cosmological data.

\subsection{Dark Matter with Long Range Forces}

The most direct approach to enhancing the clustering of matter is to increase the strength of the long range force between dark matter particles. Such long range forces are very well constrained for ordinary matter, from tests of the equivalence principle~\cite{Will:2014kxa}. However, if this new force is limited to the dark matter, it would evade most simple equivalence principle tests. It will nonetheless have observable implications for gravitational dynamics that impact structure on galactic~\cite{Kesden:2006zb,Kesden:2006vz,Keselman:2009nx,Bogorad:2023wzn} and cosmological scales~\cite{Archidiacono:2022iuu,Bottaro:2023wkd}. Interestingly, any such force would also violate the single-field consistency conditions for large-scale structure and thus would leave a measurable non-Gaussian imprint on cosmological correlators~\cite{Creminelli:2013mca,Creminelli:2013poa,Creminelli:2013nua}, in addition to the any change to the power spectrum.

Following~\cite{Saracco:2009df,Creminelli:2013nua,Bottaro:2023wkd}, suppose we introduce a massless field $\varphi$ that couples only to the dark matter with a $r^{-2}$ force similar to Newtonian gravity. This force will modify the momentum conservation equation for the dark matter,  
\beq
    \dot u_{\rm cdm}+H  u_{\rm cdm} =-\frac{1}{a} 
\left(\Phi+ \alpha \varphi\right) \ , \qquad \nabla^2 \varphi = \alpha 8 \pi G \bar \rho_{\rm cdm} \delta_{\rm cdm} \ .
\eeq
The resulting linear growth of the dark matter and baryons at $k \gg k_{\rm fs}$ where $\delta_\nu = 0$, is described by 
\begin{align}
 \ddot{\delta}_{\rm cdm}+\frac{4}{3 t} \dot{\delta}_{\rm cdm} &=\frac{2}{3 t^{2}}\left[(1-f_\nu-f_b) (1+ 2\alpha^2) \delta_{\rm cdm} + f_b \delta_{\rm b} \right] \ , \\
  \ddot{\delta}_{\rm b}+\frac{4}{3 t} \dot{\delta}_{\rm b} &=\frac{2}{3 t^{2}}\left[(1-f_\nu-f_b) \delta_{\rm cdm} + f_b \delta_{\rm b} \right] \ .
\end{align}
We will define the new growth term as $(1-f_\nu -f_b)(1+2\alpha^2) = 1+\epsilon$, so that $\epsilon$ controls the change to the linear evolution. Taking $\delta_{\rm cdm} = t^\gamma$ and $\delta_b = \xi \delta_{\rm cdm}$ we find
\beq
    \gamma(\gamma-1) + \frac{4}{3} \gamma - \frac{2}{3}( 1+ \epsilon +\xi f_b ) = 0 \ , \qquad \xi \gamma(\gamma-1) + \xi \frac{4}{3} \gamma - \frac{2}{3}\left( \frac{1+ \epsilon}{1+2\alpha^2} +\xi f_b  \right) = 0 \ .
\eeq
To linear order in $\epsilon$ and $f_b$ one finds the growing solution 
\beq
\gamma = \frac{2}{3}+ \frac{2}{5} (\epsilon + f_b) \ , \qquad \xi = 1 - (2 \alpha^2) \ .
\eeq
In the presense of this new long range force, the power spectrum is therefore modified
\beq
    P^{(\epsilon,\sum m_\nu)} (k\gg k_{\rm fs}, z ) \approx  \left(1 - 2 f_\nu  + \frac{6 }{5} (\epsilon + f_b)\log \frac{1+z_\star}{1+z} \right) P^{(\epsilon =0,\sum m_\nu=0)}(k\gg k_{\rm fs}, z ) \ .
\eeq
Here $z_\star$ is the redshift where the long-range force becomes important. In most simple models, $z_\star$ is the redshift of horizon entry $k= a(z_\star) H(z_\star)$. This would make the above signal scale dependent and thus would not mimic the neutrino signal. As a result, cosmological constraints already exclude $\alpha < 0.01$~\cite{Archidiacono:2022iuu,Bottaro:2023wkd}. Therefore, it is important that $z_\star$ is a $k$-independent constant and that the field $\varphi$ only becomes important at late times. In this case, if we assume the minimum $\sum m_\nu$ so that $f_\nu = 4 \times 10^{-3}$, as derived in Equation~(\ref{eq:fnu}), we could explain an apparent $\sum m_\nu \approx -160$ meV with $\alpha^2 = 7 \times 10^{-3}$.

A phase transition, or some other time- or temperature-dependent physics could change the mass of $\varphi$ so that it became massless at $z_\star \approx 100$. This would imply equivalence principle violation for the dark matter at later times. Current constraints~\cite{Kesden:2006zb,Kesden:2006vz,Keselman:2009nx,Bogorad:2023wzn} likely require $\alpha^2 \ll 1$ but have not been explored in detail. In addition, this type of equivalence principle violation leaves a number of cosmological~\cite{Creminelli:2013nua} and astrophysical signals~\cite{Bai:2023lyf} that could be observed in near-term surveys and experiments. For example, the change to the evolution of matter also alters the galaxy bispectrum in a way that breaks the single-field consistency conditions. This effect is sufficient to measure $\alpha^2 \gtrsim 10^{-3}$~\cite{Creminelli:2013nua} for a quasi-realistic survey. %It is plausible that these scenarios 

\subsection{Primordial Trispectrum}

The trispectrum (four-point function) of the CMB plays two significant roles in the measurement of neutrino mass. First, gravitational lensing induces a connected four point function, and measuring the trispectrum allows us to reconstruct the lensing power spectrum. Secondly, the trispectrum is also what determines the variance of the primary CMB which sets the uncertainty in all our cosmic parameters~\cite{Hu:2001fa,Okamoto:2002ik}.

A primordial trispectrum of the appropriate shape could mimic the effect of lensing and thus could lead to an apparent increase in the lensing amplitude. Both lensing and primordial trispectra can be measured using the same class of estimators defined in Ref.~\cite{Hanson:2009gu}. Concretely, we could couple the inflaton to an additional field, $\sigma(\x)$, that modulates the amplitude of the adiabatic fluctuations, $\zeta(\x)$, by a term
\beq
    \zeta(\x) = \zeta_{\rm G}(\x) + \sqrt{\tau^\sigma_{\rm NL}} \zeta_{\rm G}(\x) \sigma(\x) \ .
\eeq
where $\zeta_{\rm G}(\x)$ and $\sigma(\x)$ are Gaussian random fields. This modulation leads to a connected trispectrum
\bea
    \left\langle\zeta_{\k_1} \zeta_{\k_2} \zeta_{\k_3} \zeta_{\k_4}\right\rangle' &=& \tnl^{\sigma} P_\zeta(k_1) P_\zeta(k_3) P_\sigma(|\k_1+\k_2|) + {\rm permutations} \nonumber \\
    &\equiv &\tnl^{\sigma} T(\k_1,\k_2,\k_3,\k_4) \, .
\eea
This is not equivalent to the lensing signal because it is a three-dimensional correlation between the modes, rather than two dimensional. Bounds on this kind of non-Gaussianity for a scale invariant $\sigma$, $P_\sigma \approx P_\zeta$, have been derived from the CMB and yield $\tau^{\rm local}_{\rm NL} < 1700$ (95\%)~\cite{Marzouk:2022utf}. However, if the power spectrum of $\sigma$ were taken to be scale dependent to be degenerate with the lensing potential, $\phi(\vec L)$, it would be projected out of that analysis. Following Ref.~\cite{Smith:2015uia} (see also Refs.~\cite{Smith:2011if,Green:2023ids}), we can estimate how correlated the proposed trispectrum would be with the local model using the Fisher matrix, 
\beq
    F\left(T_1, T_2\right)=V\int \frac{d^3 \k_1 d^3 \k_2 d^3 \k_3 d^3 \k_4}{(2 \pi)^{12}} \frac{\left\langle\zeta_{\k_1} \zeta_{\k_2} \zeta_{\k_3} \zeta_{\k_4}\right\rangle_1^{\prime}\left\langle\zeta_{\k_1} \zeta_{\k_2} \zeta_{\k_3} \zeta_{\k_4}\right\rangle_2^{\prime}}{P_\zeta\left(k_1\right) P_\zeta\left(k_2\right) P_\zeta\left(k_3\right) P_\zeta\left(k_4\right)}(2 \pi)^3 \delta^3\left(\sum \k_i\right) \ ,
\eeq
where $V \propto k_{\rm min}^{-3}$ is the survey volume. The ratio of the off-diagonal to diagonal terms defines the correlation coefficient between $\tnl^{\sigma }$ and $\tnl^{\rm local}$, $C(\tnl^{\sigma },\tnl^{\rm local})$, which is approximately
\beq
    C(\tnl^{\sigma },\tnl^{\rm local})\approx \frac{\int d^3 k P_\sigma(k) P_{\zeta}(k)}{\sqrt{\int d^3 k P_\zeta(k)^2 } \sqrt{\int d^3 k P_\sigma(k)^2 }} \, .
\eeq
To match the CMB lensing power, we should choose $P_\sigma(k) \propto P_m(k)$ so that it takes a similar form to the lensing signal. We therefore require $P_\sigma(k) \to 0$ as $k\to 0$, $P_\sigma \propto k^{-3}$ as $k\to \infty$, and have a maximum at some $k=k_\star$. We would then expect the correlation to be suppressed by $C(\tnl^{\sigma },\tnl^{\rm local}) \approx (k_{\rm min} / k_\star)^3 \ll 1$. In this regard, the shape of $P_\sigma(k)$ may not have to be finely tuned to contribute to the observed lensing trispectrum without violating other CMB trispectrum constraints. Other trispectrum shapes, like those considered in Refs.~\cite{Smith:2015uia,Planck:2019kim} are usually scale invariant and peak in equilateral configurations where $k \sim k_{\rm max}$.

Although this signal would be degenerate with lensing in the CMB, it would be introduce non-Gaussianity in the late universe that could be measured through the galaxy power spectrum~\cite{Ferraro:2014jba} (via scale-dependent bias~\cite{Gong:2011gx,Baumann:2012bc}) or cross-correlations between the CMB and LSS~\cite{AnilKumar:2022flx}. CMB lensing is currently measured at $40\sigma$~\cite{Planck:2018lbu,Carron:2022eyg,ACT:2023dou,ACT:2023kun} and therefore a trispectrum mimicking a $2.5\%$--$7.5\%$ shift in the lensing amplitude would visible at the 1--3$\sigma$ level. Given that the current constraints on primordial non-Gaussianity from related models are at least an order of magnitude weaker than Planck constraints~\cite{Green:2023uyz}, we do not expect\footnote{We are not aware of published constraints on $\tau_{\rm NL}$ from current galaxy survey data in which we can directly compare Planck.} current galaxy survey data to be sensitive to such a trispectrum. However, data from DESI, Euclid~\cite{EUCLID:2011zbd}, and particularly SPHEREx~\cite{SPHEREx:2014bgr} are expected to be up to an order of magnitude more sensitive than Planck to this type of non-Gaussian signature. Concretely, SPHEREx is expected to be sensitive to $\tau_{\rm NL} = 130$ at $2\sigma$~\cite{Shiveshwarkar:2023afl} which is roughly 10 times the sensitivity of Planck~\cite{Planck:2019kim}.

A second possibility is that additional contributions to the trispectrum could increase the true uncertainty in cosmic parameters. This could increase the probability that value of $A_s$ determined from the primary CMB is simply a statistical outlier. Specifically, a large primordial trispectrum increases the deviation of parameters from their mean values. One model that achieves such behavior is disorder in single field inflation~\cite{Green:2014xqa}. In these models, random features in the inflationary potential introduce, on average, a trispectrum that is identical the Gaussian noise but with a larger or smaller amplitude. One can achieve a similar effect on $\sum m_\nu$~\cite{Forconi:2023akg} from super-sample covariance~\cite{Takada:2013wfa}, through a large amplitude of local-term non-Gaussianity (e.g.~$\tnl^{\rm local}$). To be consistent with CMB constraints, the effective amplitude $\tnl$ would have to be scale-dependent to avoid the direct constraints from the CMB trispectrum.

%%%%%%%%%%%%%%%%%%%%%%%%%%%%%%%%%%%%%%%
\section{Conclusions}\label{sec:conclusions}
%%%%%%%%%%%%%%%%%%%%%%%%%%%%%%%%%%%%%%%

The exclusion of the minimum sum of neutrino masses, from either the inverted or normal hierarchy, is a remarkable statement of the power of cosmological data. At these masses, neutrinos form only a fraction of a percent of the total energy density of the universe. The presence of cosmic neutrinos has been robustly established during the era of nucleosynthesis~\cite{Fields:2019pfx} (BBN) and recombination~\cite{Planck:2018vyg} (CMB), through the measurement of $\Neff$ and therefore their small but measurable impact on the late universe was to be expected. As we have no simple path to a direct measurement of cosmic neutrinos on earth, cosmological observations provide a novel window into the universe, capable of revealing new secrets.

The recent BAO measurements from DESI enrich this story. Allowing $\sum m_\nu < 0$, as an indication of enhanced of clustering, we find data from CMB+DESI constrains $\sum m_\nu = - 160 \pm 90$ meV (68\%), excluding at about 3$\sigma$ even the minimum neutrino masses consistent with neutrino oscillation experiments. Yet, we showed that this measurement can be naturally explained by new physics in the neutrino and/or dark sectors that is otherwise weakly constrained by other experiments and observations. A measurement consistent with $\sum m_\nu = 0$ could be naturally explained by neutrino decays, cooling, or time-dependent neutrino masses, pointing to new physics coupled to neutrinos and potentially dark matter (sectors). Achieving $\sum m_\nu < 0$ requires physics beyond the neutrino sector but could be explained by new long range forces for dark matter or changes to the primordial statistics. Each class of models naturally suggests signals that could be present in existing data or testable with near term experiments or observations.

It is important that the measurement of $\sum m_\nu$ from the CMB and DESI is incompatible with a wide range of proposals for BSM physics that are also otherwise unconstrained. Light but massive relics~\cite{Xu:2021rwg} are extremely common in models of BSM physics, including many approaches to the hierarchy problem, explanations of dark matter, models including light gravitinos~\cite{Moroi:1993mb,Osato:2016ixc}, etc. These necessarily contribute positively to $\Neff$ and $\sum m_\nu$ and thus would further exacerbate the tension with the minimum sum of neutrino masses. As a result, any such model would have to incorporate additional physics, of the kind discussed in this paper, in addition to the new physics relevant to these problems. It is interesting that our results from neutrino decay point to a possible origin from new physics at 10-100 TeV, which could provide a common origin for both effects.

\paragraph{Acknowledgements}
We are grateful to Kim Berghaus, Tim Cohen, Raphael Flauger, George Fuller, Peter Graham, Jiashu Han, Colin Hill, Mustapha Ishak, Thomas Konstandin, Tongyan Lin, and Ben Wallisch for helpful discussions. NC is supported by the US~Department of Energy under grant \mbox{DE-SC0011702}. DG is supported by the US~Department of Energy under grant~\mbox{DE-SC0009919}. This work was supported by the U.S.~Department of Energy~(DOE), Office of Science, National Quantum Information Science Research Centers, Superconducting Quantum Materials and Systems Center~(SQMS) under Contract No.~DE-AC02-07CH11359. 
S.R. is also supported in part by the U.S. National Science Foundation (NSF) under Grant
No. PHY-1818899,  the Simons Investigator Grant No. 827042,
and by the DOE under a QuantISED grant for MAGIS and Fermilab. JM is supported by the US~Department of Energy under grant~\mbox{DE-SC0010129}. Computational resources for this research were provided by SMU’s Center for Research Computing.  We acknowledge the use of \texttt{CAMB}~\cite{Lewis:1999bs}, \texttt{CLASS}~\cite{Blas:2011rf}, \texttt{IPython}~\cite{Perez:2007ipy}, 
and the Python packages \texttt{Matplotlib}~\cite{Hunter:2007mat}, \texttt{NumPy}~\cite{Harris:2020xlr}, and~\texttt{SciPy}~\cite{Virtanen:2019joe}.

\newpage
% -----------------------------------------------------------------------------------------------------------------------------------------
\appendix

%%%%%%%%%%%%%%%%%

\section{The Suppression of Clustering }\label{app:A}

In this appendix, we review the calculation of the linear growth of structure in a universe with massive neutrinos. This calculation explains the suppression of small scale power due to neutrino free streaming, which is the dominant cosmological signal responsible for the constraints on $\sum m_\nu$.

Following~\cite{Green:2022bre}, we define the density contrasts of the dark matter and baryons as 
 $\delta_\mathrm{cb}= \frac{\delta\rho_\mathrm{cdm} + \delta\rho_\mathrm{b}}{\bar{\rho}_\mathrm{cdm}+\bar{\rho}_\mathrm{b}}$, and the neutrinos, $\delta_\nu = \frac{\delta \rho_\nu}{\bar{\rho}_\nu}$. Energy and momentum conservation of these species after recombination is then described by the coupled equations
\begin{equation}
\dot \delta_{\rm cb}(\k,t)-a^{-1} k^2 u_{\rm cb}=0 \ , \qquad
\dot \delta_{\nu}(\k,t)-a^{-1} k^2 u_{\nu}=0 \ ,
\end{equation}
and
\begin{align}
\dot u_{\rm cb}+H  u_{\rm cb} =-\frac{1}{a} \Phi \ , \qquad 
\dot u_{\nu}+H u_{\nu} =-\frac{1}{a} \Phi-\frac{c_{\nu}^{2}}{a } \delta_{\nu} \ .
\end{align}
Here we have defined the scalar velocity potential $u_i$ for each species as $\vec v_i = \vec \nabla u_i$. Finally, $\Phi$ is the Newtonian gravitational potential, which obeys
\beq
\nabla^2 \Phi =  4 \pi G \left(\bar \rho_{cb} \delta_{cb} +\bar \rho_\nu \delta_\nu \right) \ .
\eeq
In a matter dominated universe, $H^2 \propto a^{-3}$ which implies that $a(t) \propto t^{2/3}$ and $\bar \rho_{\rm m} \propto t^{-2}$. Differentiating these equations allows us to eliminate the velocity potential to find two second-order equations
\begin{align}
    \ddot{\delta}_{\rm cb}+\frac{4}{3 t} \dot{\delta}_{\rm cb}&=\frac{2}{3 t^{2}}\left[f_\nu \delta_{\nu}+(1-f_\nu) \delta_{\rm cb}\right] \ , \label{eq:nu_cmd1}\\
    \ddot{\delta}_{\nu}+\frac{4}{3 t} \dot{\delta}_{\nu}&=-\frac{2 \alpha}{3 t^{2}} \delta_{\nu}+\frac{2}{3 t^{2}}\left[f_\nu \delta_{\nu}+(1-f_\nu) \delta_{\rm cb}\right] \ ,
\end{align}
where
\beq
    \alpha \equiv \frac{3 k^{2} c_\nu^{2} t^{2}}{2 a^{2}}=\frac{k^2}{k_{\rm fs}^2} \ , 
    \quad c_\nu \equiv \frac{\langle p_\nu \rangle}{m_\nu}  \ ,
    \quad f_\nu \equiv \frac{\Omega_{\nu}}{\Omega_\mathrm{m}}  \ .
\eeq
From here, one can solve these equations numerically to understand the influence of the neutrinos on the matter fluctuations in the linear regime.

In the regime $\alpha \gg 1$, it easy to understand the solutions as follows: the homogeneous equation for $\delta_\nu$ (i.e.~$\delta_{\rm cb} \approx 0$) can be solved to find that $\delta_\nu \propto t^{-1/6} \to 0$ as $t\to \infty$. We can also solve the inhomogeneous equation with $\delta_\nu = \xi \delta_{\rm cb}$ to find $\xi \propto 1/\alpha \to 0$. Therefore we can focus on $\delta_{\rm cb}$ with $\delta_\nu =0$. Taking the ansatz $\delta_{\rm cb} = t^\gamma$ and $\delta_\nu= 0$, we get
\beq
\gamma(\gamma-1) + \frac{4}{3} \gamma -\frac{2(1-f_\nu)}{3} = 0 \to \gamma = \frac{2}{3} - \frac{2}{5} f_{\nu} + {\cal O}(f_\nu^2)
\eeq
where we kept only the growing solution with $\gamma > 0$. In a matter-dominated universe, $H^2 \propto a^{-3}$ which implies that $a(t) \propto t^{2/3}$, and therefore 
\beq
\delta_{\rm cb}(\k,t)\approx \delta_{\rm cb}(\k,t_\nu)a(t)^{1- 3f_\nu/5} \ ,
\eeq
where $1+z_\nu = a(t_\nu)^{-1}$. Finally, since $\bar \rho_m =
\bar \rho_{\rm cb} +\bar \rho_\nu$, the total matter density contrast $\delta_m =\delta \rho_m /\bar \rho_m$ is given by
\bea
\delta_m (\k,t) = \frac{\delta \rho_{\rm cb} + \delta \rho_\nu}{\bar\rho_m} &=& \delta_{\rm cb}(\k,t) \approx (1-f_\nu)  \delta_{\rm cb}(\k,t_\nu)a(t)^{1- 3f_\nu/5} \nonumber \\
&\approx& \delta_{\rm cb} \left(1-f_\nu -\frac{3}{5} f_\nu \log \frac{1+z_\nu}{1+z} \right) \ .
\eea
This gives rise to the suppression of the power spectrum
\beq
    P^{(\sum m_\nu)} (k\gg k_{\rm fs}, z ) \approx  \left(1 - 2 f_\nu  -\frac{6 }{5} f_\nu\log \frac{1+z_\nu}{1+z} \right) P^{(\sum m_\nu =0)}(k\gg k_{\rm fs}, z ) \ .
\eeq
In this regard, we see that the suppression is a straightforward consequence of the linear evolution.

\clearpage
\phantomsection
\addcontentsline{toc}{section}{References}
\bibliographystyle{utphys}
\bibliography{Refs}

\providecommand{\href}[2]{#2}\begingroup\raggedright\begin{thebibliography}{100}

\bibitem{Lesgourgues:2006nd}
J.~Lesgourgues and S.~Pastor, ``{Massive neutrinos and cosmology},''
  \href{http://dx.doi.org/10.1016/j.physrep.2006.04.001}{{\em Phys. Rept.}
  {\bfseries 429} (2006) 307--379},
  \href{http://arxiv.org/abs/astro-ph/0603494}{{\ttfamily
  arXiv:astro-ph/0603494}}.

\bibitem{TopicalConvenersKNAbazajianJECarlstromATLee:2013bxd}
{\bfseries Topical Conveners: K.N. Abazajian, J.E. Carlstrom, A.T. Lee}
  Collaboration, K.~N. Abazajian {\em et~al.}, ``{Neutrino Physics from the
  Cosmic Microwave Background and Large Scale Structure},''
  \href{http://dx.doi.org/10.1016/j.astropartphys.2014.05.014}{{\em Astropart.
  Phys.} {\bfseries 63} (2015) 66--80},
  \href{http://arxiv.org/abs/1309.5383}{{\ttfamily arXiv:1309.5383
  [astro-ph.CO]}}.

\bibitem{Dvorkin:2019jgs}
C.~Dvorkin {\em et~al.}, ``{Neutrino Mass from Cosmology: Probing Physics
  Beyond the Standard Model},''
  \href{http://arxiv.org/abs/1903.03689}{{\ttfamily arXiv:1903.03689
  [astro-ph.CO]}}.

\bibitem{ParticleDataGroup:2020ssz}
{\bfseries Particle Data Group} Collaboration, P.~A. Zyla {\em et~al.},
  ``{Review of Particle Physics},''
  \href{http://dx.doi.org/10.1093/ptep/ptaa104}{{\em PTEP} {\bfseries 2020}
  no.~8, (2020) 083C01}.

\bibitem{Font-Ribera:2013rwa}
A.~Font-Ribera, P.~McDonald, N.~Mostek, B.~A. Reid, H.-J. Seo, and A.~Slosar,
  ``{DESI and other dark energy experiments in the era of neutrino mass
  measurements},'' \href{http://dx.doi.org/10.1088/1475-7516/2014/05/023}{{\em
  JCAP} {\bfseries 05} (2014) 023},
  \href{http://arxiv.org/abs/1308.4164}{{\ttfamily arXiv:1308.4164
  [astro-ph.CO]}}.

\bibitem{CMB-S4:2016ple}
{\bfseries CMB-S4} Collaboration, K.~N. Abazajian {\em et~al.}, ``{CMB-S4
  Science Book, First Edition},''
  \href{http://arxiv.org/abs/1610.02743}{{\ttfamily arXiv:1610.02743
  [astro-ph.CO]}}.

\bibitem{DESI:2016fyo}
{\bfseries DESI} Collaboration, A.~Aghamousa {\em et~al.}, ``{The DESI
  Experiment Part I: Science,Targeting, and Survey Design},''
  \href{http://arxiv.org/abs/1611.00036}{{\ttfamily arXiv:1611.00036
  [astro-ph.IM]}}.

\bibitem{Kaplinghat:2003bh}
M.~Kaplinghat, L.~Knox, and Y.-S. Song, ``{Determining neutrino mass from the
  CMB alone},'' \href{http://dx.doi.org/10.1103/PhysRevLett.91.241301}{{\em
  Phys. Rev. Lett.} {\bfseries 91} (2003) 241301},
  \href{http://arxiv.org/abs/astro-ph/0303344}{{\ttfamily
  arXiv:astro-ph/0303344}}.

\bibitem{Pan:2015bgi}
Z.~Pan and L.~Knox, ``{Constraints on neutrino mass from Cosmic Microwave
  Background and Large Scale Structure},''
  \href{http://dx.doi.org/10.1093/mnras/stv2164}{{\em Mon. Not. Roy. Astron.
  Soc.} {\bfseries 454} no.~3, (2015) 3200--3206},
  \href{http://arxiv.org/abs/1506.07493}{{\ttfamily arXiv:1506.07493
  [astro-ph.CO]}}.

\bibitem{DESI:2024mwx}
{\bfseries DESI} Collaboration, A.~G. Adame {\em et~al.}, ``{DESI 2024 VI:
  Cosmological Constraints from the Measurements of Baryon Acoustic
  Oscillations},'' \href{http://arxiv.org/abs/2404.03002}{{\ttfamily
  arXiv:2404.03002 [astro-ph.CO]}}.

\bibitem{Aghanim:2019ame}
{\bfseries Planck} Collaboration, N.~Aghanim {\em et~al.}, ``{Planck 2018
  results. V. CMB power spectra and likelihoods},''
  \href{http://dx.doi.org/10.1051/0004-6361/201936386}{{\em Astron. Astrophys.}
  {\bfseries 641} (2020) A5}, \href{http://arxiv.org/abs/1907.12875}{{\ttfamily
  arXiv:1907.12875 [astro-ph.CO]}}.

\bibitem{Carron:2022eyg}
J.~Carron, M.~Mirmelstein, and A.~Lewis, ``{CMB lensing from Planck
  PR4~maps},'' \href{http://dx.doi.org/10.1088/1475-7516/2022/09/039}{{\em
  JCAP} {\bfseries 09} (2022) 039},
  \href{http://arxiv.org/abs/2206.07773}{{\ttfamily arXiv:2206.07773
  [astro-ph.CO]}}.

\bibitem{ACT:2023dou}
{\bfseries ACT} Collaboration, F.~J. Qu {\em et~al.}, ``{The Atacama Cosmology
  Telescope: A Measurement of the DR6 CMB Lensing Power Spectrum and Its
  Implications for Structure Growth},''
  \href{http://dx.doi.org/10.3847/1538-4357/acfe06}{{\em Astrophys. J.}
  {\bfseries 962} no.~2, (2024) 112},
  \href{http://arxiv.org/abs/2304.05202}{{\ttfamily arXiv:2304.05202
  [astro-ph.CO]}}.

\bibitem{ACT:2023kun}
{\bfseries ACT} Collaboration, M.~S. Madhavacheril {\em et~al.}, ``{The Atacama
  Cosmology Telescope: DR6 Gravitational Lensing Map and Cosmological
  Parameters},'' \href{http://dx.doi.org/10.3847/1538-4357/acff5f}{{\em
  Astrophys. J.} {\bfseries 962} no.~2, (2024) 113},
  \href{http://arxiv.org/abs/2304.05203}{{\ttfamily arXiv:2304.05203
  [astro-ph.CO]}}.

\bibitem{Brieden:2022lsd}
S.~Brieden, H.~Gil-Mar\'\i{}n, and L.~Verde, ``{Model-agnostic interpretation
  of 10 billion years of cosmic evolution traced by BOSS and eBOSS data},''
  \href{http://dx.doi.org/10.1088/1475-7516/2022/08/024}{{\em JCAP} {\bfseries
  08} no.~08, (2022) 024}, \href{http://arxiv.org/abs/2204.11868}{{\ttfamily
  arXiv:2204.11868 [astro-ph.CO]}}.

\bibitem{Palanque-Delabrouille:2019iyz}
N.~Palanque-Delabrouille, C.~Y\`eche, N.~Sch\"oneberg, J.~Lesgourgues,
  M.~Walther, S.~Chabanier, and E.~Armengaud, ``{Hints, neutrino bounds and WDM
  constraints from SDSS DR14 Lyman-$\alpha$ and Planck full-survey data},''
  \href{http://dx.doi.org/10.1088/1475-7516/2020/04/038}{{\em JCAP} {\bfseries
  04} (2020) 038}, \href{http://arxiv.org/abs/1911.09073}{{\ttfamily
  arXiv:1911.09073 [astro-ph.CO]}}.

\bibitem{Planck:2018vyg}
{\bfseries Planck} Collaboration, N.~Aghanim {\em et~al.}, ``{Planck 2018
  results. VI. Cosmological parameters},''
  \href{http://dx.doi.org/10.1051/0004-6361/201833910}{{\em Astron. Astrophys.}
  {\bfseries 641} (2020) A6}, \href{http://arxiv.org/abs/1807.06209}{{\ttfamily
  arXiv:1807.06209 [astro-ph.CO]}}. [Erratum: Astron.Astrophys. 652, C4
  (2021)].

\bibitem{Follin:2015hya}
B.~Follin, L.~Knox, M.~Millea, and Z.~Pan, ``{First Detection of the Acoustic
  Oscillation Phase Shift Expected from the Cosmic Neutrino Background},''
  \href{http://dx.doi.org/10.1103/PhysRevLett.115.091301}{{\em Phys. Rev.
  Lett.} {\bfseries 115} no.~9, (2015) 091301},
  \href{http://arxiv.org/abs/1503.07863}{{\ttfamily arXiv:1503.07863
  [astro-ph.CO]}}.

\bibitem{Baumann:2015rya}
D.~Baumann, D.~Green, J.~Meyers, and B.~Wallisch, ``{Phases of New Physics in
  the CMB},'' \href{http://dx.doi.org/10.1088/1475-7516/2016/01/007}{{\em JCAP}
  {\bfseries 01} (2016) 007}, \href{http://arxiv.org/abs/1508.06342}{{\ttfamily
  arXiv:1508.06342 [astro-ph.CO]}}.

\bibitem{Baumann:2019keh}
D.~Baumann, F.~Beutler, R.~Flauger, D.~Green, A.~Slosar, M.~Vargas-Maga\~na,
  B.~Wallisch, and C.~Y\`eche, ``{First constraint on the neutrino-induced
  phase shift in the spectrum of baryon acoustic oscillations},''
  \href{http://dx.doi.org/10.1038/s41567-019-0435-6}{{\em Nature Phys.}
  {\bfseries 15} (2019) 465--469},
  \href{http://arxiv.org/abs/1803.10741}{{\ttfamily arXiv:1803.10741
  [astro-ph.CO]}}.

\bibitem{Cyr-Racine:2013jua}
F.-Y. Cyr-Racine and K.~Sigurdson, ``{Limits on Neutrino-Neutrino Scattering in
  the Early Universe},''
  \href{http://dx.doi.org/10.1103/PhysRevD.90.123533}{{\em Phys. Rev. D}
  {\bfseries 90} no.~12, (2014) 123533},
  \href{http://arxiv.org/abs/1306.1536}{{\ttfamily arXiv:1306.1536
  [astro-ph.CO]}}.

\bibitem{Lancaster:2017ksf}
L.~Lancaster, F.-Y. Cyr-Racine, L.~Knox, and Z.~Pan, ``{A tale of two modes:
  Neutrino free-streaming in the early universe},''
  \href{http://dx.doi.org/10.1088/1475-7516/2017/07/033}{{\em JCAP} {\bfseries
  07} (2017) 033}, \href{http://arxiv.org/abs/1704.06657}{{\ttfamily
  arXiv:1704.06657 [astro-ph.CO]}}.

\bibitem{He:2023oke}
A.~He, R.~An, M.~M. Ivanov, and V.~Gluscevic, ``{Self-Interacting Neutrinos in
  Light of Large-Scale Structure Data},''
  \href{http://arxiv.org/abs/2309.03956}{{\ttfamily arXiv:2309.03956
  [astro-ph.CO]}}.

\bibitem{Camarena:2023cku}
D.~Camarena, F.-Y. Cyr-Racine, and J.~Houghteling, ``{Confronting
  self-interacting neutrinos with the full shape of the galaxy power
  spectrum},'' \href{http://dx.doi.org/10.1103/PhysRevD.108.103535}{{\em Phys.
  Rev. D} {\bfseries 108} no.~10, (2023) 103535},
  \href{http://arxiv.org/abs/2309.03941}{{\ttfamily arXiv:2309.03941
  [astro-ph.CO]}}.

\bibitem{Camarena:2024zck}
D.~Camarena and F.-Y. Cyr-Racine, ``{Absence of concordance in a simple
  self-interacting neutrino cosmology},''
  \href{http://arxiv.org/abs/2403.05496}{{\ttfamily arXiv:2403.05496
  [astro-ph.CO]}}.

\bibitem{Green:2021xzn}
D.~Green and J.~Meyers, ``{Cosmological Implications of a Neutrino Mass
  Detection},'' \href{http://arxiv.org/abs/2111.01096}{{\ttfamily
  arXiv:2111.01096 [astro-ph.CO]}}.

\bibitem{Green:2022bre}
D.~Green, ``{Cosmic Signals of Fundamental Physics},''
  \href{http://dx.doi.org/10.22323/1.439.0005}{{\em PoS} {\bfseries TASI2022}
  (2024) 005}, \href{http://arxiv.org/abs/2212.08685}{{\ttfamily
  arXiv:2212.08685 [hep-ph]}}.

\bibitem{Lewis:2006fu}
A.~Lewis and A.~Challinor, ``{Weak gravitational lensing of the CMB},''
  \href{http://dx.doi.org/10.1016/j.physrep.2006.03.002}{{\em Phys. Rept.}
  {\bfseries 429} (2006) 1--65},
  \href{http://arxiv.org/abs/astro-ph/0601594}{{\ttfamily
  arXiv:astro-ph/0601594}}.

\bibitem{eBOSS:2020yzd}
{\bfseries eBOSS} Collaboration, S.~Alam {\em et~al.}, ``{Completed SDSS-IV
  extended Baryon Oscillation Spectroscopic Survey: Cosmological implications
  from two decades of spectroscopic surveys at the Apache Point Observatory},''
  \href{http://dx.doi.org/10.1103/PhysRevD.103.083533}{{\em Phys. Rev. D}
  {\bfseries 103} no.~8, (2021) 083533},
  \href{http://arxiv.org/abs/2007.08991}{{\ttfamily arXiv:2007.08991
  [astro-ph.CO]}}.

\bibitem{Lewis:1999bs}
A.~Lewis, A.~Challinor, and A.~Lasenby, ``{Efficient Computation of CMB
  Anisotropies in Closed FRW Models},''
  \href{http://dx.doi.org/10.1086/309179}{{\em Astrophys. J.} {\bfseries 538}
  (2000) 473}, \href{http://arxiv.org/abs/astro-ph/9911177}{{\ttfamily
  arXiv:astro-ph/9911177}}.

\bibitem{Howlett:2012mh}
C.~Howlett, A.~Lewis, A.~Hall, and A.~Challinor, ``{CMB power spectrum
  parameter degeneracies in the era of precision cosmology},''
  \href{http://dx.doi.org/10.1088/1475-7516/2012/04/027}{{\em JCAP} {\bfseries
  1204} (2012) 027}, \href{http://arxiv.org/abs/1201.3654}{{\ttfamily
  arXiv:1201.3654 [astro-ph.CO]}}.
\url{https://arxiv.org/abs/1201.3654}.
%%CITATION = ARXIV:1201.3654;%%.

\bibitem{Blas:2011rf}
D.~Blas, J.~Lesgourgues, and T.~Tram, ``{The Cosmic Linear Anisotropy Solving
  System~(CLASS)~II: Approximation Schemes},''
  \href{http://dx.doi.org/10.1088/1475-7516/2011/07/034}{{\em JCAP} {\bfseries
  07} (2011) 034}, \href{http://arxiv.org/abs/1104.2933}{{\ttfamily
  arXiv:1104.2933 [astro-ph.CO]}}.

\bibitem{DESI:2024lzq}
{\bfseries DESI} Collaboration, A.~G. Adame {\em et~al.}, ``{DESI 2024 IV:
  Baryon Acoustic Oscillations from the Lyman Alpha Forest},''
  \href{http://arxiv.org/abs/2404.03001}{{\ttfamily arXiv:2404.03001
  [astro-ph.CO]}}.

\bibitem{DESI:2024uvr}
{\bfseries DESI} Collaboration, A.~G. Adame {\em et~al.}, ``{DESI 2024 III:
  Baryon Acoustic Oscillations from Galaxies and Quasars},''
  \href{http://arxiv.org/abs/2404.03000}{{\ttfamily arXiv:2404.03000
  [astro-ph.CO]}}.

\bibitem{Torrado:2020dgo}
J.~Torrado and A.~Lewis, ``{Cobaya: Code for Bayesian Analysis of hierarchical
  physical models},''
  \href{http://dx.doi.org/10.1088/1475-7516/2021/05/057}{{\em JCAP} {\bfseries
  05} (2021) 057}, \href{http://arxiv.org/abs/2005.05290}{{\ttfamily
  arXiv:2005.05290 [astro-ph.IM]}}.

\bibitem{Lewis:2002ah}
A.~Lewis and S.~Bridle, ``{Cosmological parameters from CMB and other data: A
  Monte Carlo approach},''
  \href{http://dx.doi.org/10.1103/PhysRevD.66.103511}{{\em Phys. Rev.}
  {\bfseries D66} (2002) 103511},
  \href{http://arxiv.org/abs/astro-ph/0205436}{{\ttfamily
  arXiv:astro-ph/0205436 [astro-ph]}}.
\url{https://arxiv.org/abs/astro-ph/0205436}.
%%CITATION = ASTRO-PH/0205436;%%.

\bibitem{Lewis:2013hha}
A.~Lewis, ``{Efficient sampling of fast and slow cosmological parameters},''
  \href{http://dx.doi.org/10.1103/PhysRevD.87.103529}{{\em Phys. Rev.}
  {\bfseries D87} no.~10, (2013) 103529},
  \href{http://arxiv.org/abs/1304.4473}{{\ttfamily arXiv:1304.4473
  [astro-ph.CO]}}.
\url{https://arxiv.org/abs/1304.4473}.
%%CITATION = ARXIV:1304.4473;%%.

\bibitem{Neal:2005}
R.~M. {Neal}, ``{Taking Bigger Metropolis Steps by Dragging Fast Variables},''
  {\em ArXiv Mathematics e-prints} (Feb., 2005) ,
  \href{http://arxiv.org/abs/math/0502099}{{\ttfamily math/0502099}}.
  \url{https://arxiv.org/abs/math/0502099}.

\bibitem{Abdalla:2022yfr}
E.~Abdalla {\em et~al.}, ``{Cosmology intertwined: A review of the particle
  physics, astrophysics, and cosmology associated with the cosmological
  tensions and anomalies},''
  \href{http://dx.doi.org/10.1016/j.jheap.2022.04.002}{{\em JHEAp} {\bfseries
  34} (2022) 49--211}, \href{http://arxiv.org/abs/2203.06142}{{\ttfamily
  arXiv:2203.06142 [astro-ph.CO]}}.

\bibitem{WMAP:2003ivt}
{\bfseries WMAP} Collaboration, C.~L. Bennett {\em et~al.}, ``{First year
  Wilkinson Microwave Anisotropy Probe (WMAP) observations: Preliminary maps
  and basic results},'' \href{http://dx.doi.org/10.1086/377253}{{\em Astrophys.
  J. Suppl.} {\bfseries 148} (2003) 1--27},
  \href{http://arxiv.org/abs/astro-ph/0302207}{{\ttfamily
  arXiv:astro-ph/0302207}}.

\bibitem{WMAP:2006bqn}
{\bfseries WMAP} Collaboration, D.~N. Spergel {\em et~al.}, ``{Wilkinson
  Microwave Anisotropy Probe (WMAP) three year results: implications for
  cosmology},'' \href{http://dx.doi.org/10.1086/513700}{{\em Astrophys. J.
  Suppl.} {\bfseries 170} (2007) 377},
  \href{http://arxiv.org/abs/astro-ph/0603449}{{\ttfamily
  arXiv:astro-ph/0603449}}.

\bibitem{WMAP:2008lyn}
{\bfseries WMAP} Collaboration, E.~Komatsu {\em et~al.}, ``{Five-Year Wilkinson
  Microwave Anisotropy Probe (WMAP) Observations: Cosmological
  Interpretation},'' \href{http://dx.doi.org/10.1088/0067-0049/180/2/330}{{\em
  Astrophys. J. Suppl.} {\bfseries 180} (2009) 330--376},
  \href{http://arxiv.org/abs/0803.0547}{{\ttfamily arXiv:0803.0547
  [astro-ph]}}.

\bibitem{WMAP:2010qai}
{\bfseries WMAP} Collaboration, E.~Komatsu {\em et~al.}, ``{Seven-Year
  Wilkinson Microwave Anisotropy Probe (WMAP) Observations: Cosmological
  Interpretation},'' \href{http://dx.doi.org/10.1088/0067-0049/192/2/18}{{\em
  Astrophys. J. Suppl.} {\bfseries 192} (2011) 18},
  \href{http://arxiv.org/abs/1001.4538}{{\ttfamily arXiv:1001.4538
  [astro-ph.CO]}}.

\bibitem{WMAP:2012nax}
{\bfseries WMAP} Collaboration, G.~Hinshaw {\em et~al.}, ``{Nine-Year Wilkinson
  Microwave Anisotropy Probe (WMAP) Observations: Cosmological Parameter
  Results},'' \href{http://dx.doi.org/10.1088/0067-0049/208/2/19}{{\em
  Astrophys. J. Suppl.} {\bfseries 208} (2013) 19},
  \href{http://arxiv.org/abs/1212.5226}{{\ttfamily arXiv:1212.5226
  [astro-ph.CO]}}.

\bibitem{Planck:2013pxb}
{\bfseries Planck} Collaboration, P.~A.~R. Ade {\em et~al.}, ``{Planck 2013
  results. XVI. Cosmological parameters},''
  \href{http://dx.doi.org/10.1051/0004-6361/201321591}{{\em Astron. Astrophys.}
  {\bfseries 571} (2014) A16}, \href{http://arxiv.org/abs/1303.5076}{{\ttfamily
  arXiv:1303.5076 [astro-ph.CO]}}.

\bibitem{Planck:2015fie}
{\bfseries Planck} Collaboration, P.~A.~R. Ade {\em et~al.}, ``{Planck 2015
  results. XIII. Cosmological parameters},''
  \href{http://dx.doi.org/10.1051/0004-6361/201525830}{{\em Astron. Astrophys.}
  {\bfseries 594} (2016) A13},
  \href{http://arxiv.org/abs/1502.01589}{{\ttfamily arXiv:1502.01589
  [astro-ph.CO]}}.

\bibitem{Planck:2020olo}
{\bfseries Planck} Collaboration, Y.~Akrami {\em et~al.}, ``{$Planck$
  intermediate results. LVII. Joint Planck LFI and HFI data processing},''
  \href{http://dx.doi.org/10.1051/0004-6361/202038073}{{\em Astron. Astrophys.}
  {\bfseries 643} (2020) A42},
  \href{http://arxiv.org/abs/2007.04997}{{\ttfamily arXiv:2007.04997
  [astro-ph.CO]}}.

\bibitem{Tristram:2023haj}
M.~Tristram {\em et~al.}, ``{Cosmological parameters derived from the final
  Planck data release (PR4)},''
  \href{http://dx.doi.org/10.1051/0004-6361/202348015}{{\em Astron. Astrophys.}
  {\bfseries 682} (2024) A37},
  \href{http://arxiv.org/abs/2309.10034}{{\ttfamily arXiv:2309.10034
  [astro-ph.CO]}}.

\bibitem{Essinger-Hileman:2014pja}
T.~Essinger-Hileman {\em et~al.}, ``{CLASS: The Cosmology Large Angular Scale
  Surveyor},'' \href{http://dx.doi.org/10.1117/12.2056701}{{\em Proc. SPIE Int.
  Soc. Opt. Eng.} {\bfseries 9153} (2014) 91531I},
  \href{http://arxiv.org/abs/1408.4788}{{\ttfamily arXiv:1408.4788
  [astro-ph.IM]}}.

\bibitem{Eimer:2023esh}
J.~R. Eimer {\em et~al.}, ``{CLASS Angular Power Spectra and Map-component
  Analysis for 40 GHz Observations through 2022},''
  \href{http://dx.doi.org/10.3847/1538-4357/ad1abf}{{\em Astrophys. J.}
  {\bfseries 963} no.~2, (2024) 92},
  \href{http://arxiv.org/abs/2309.00675}{{\ttfamily arXiv:2309.00675
  [astro-ph.CO]}}.

\bibitem{LiteBIRD:2022cnt}
{\bfseries LiteBIRD} Collaboration, E.~Allys {\em et~al.}, ``{Probing Cosmic
  Inflation with the LiteBIRD Cosmic Microwave Background Polarization
  Survey},'' \href{http://dx.doi.org/10.1093/ptep/ptac150}{{\em PTEP}
  {\bfseries 2023} no.~4, (2023) 042F01},
  \href{http://arxiv.org/abs/2202.02773}{{\ttfamily arXiv:2202.02773
  [astro-ph.IM]}}.

\bibitem{Errard:2022fcm}
J.~Errard, M.~Remazeilles, J.~Aumont, J.~Delabrouille, D.~Green, S.~Hanany,
  B.~S. Hensley, and A.~Kogut, ``{Constraints on the Optical Depth to
  Reionization from Balloon-borne Cosmic Microwave Background Measurements},''
  \href{http://dx.doi.org/10.3847/1538-4357/ac9978}{{\em Astrophys. J.}
  {\bfseries 940} no.~1, (2022) 68},
  \href{http://arxiv.org/abs/2206.03389}{{\ttfamily arXiv:2206.03389
  [astro-ph.CO]}}.

\bibitem{Yu:2018tem}
B.~Yu, R.~Z. Knight, B.~D. Sherwin, S.~Ferraro, L.~Knox, and M.~Schmittfull,
  ``{Toward neutrino mass from cosmology without optical depth information},''
  \href{http://dx.doi.org/10.1103/PhysRevD.107.123522}{{\em Phys. Rev. D}
  {\bfseries 107} no.~12, (2023) 123522},
  \href{http://arxiv.org/abs/1809.02120}{{\ttfamily arXiv:1809.02120
  [astro-ph.CO]}}.

\bibitem{Brinckmann:2018owf}
T.~Brinckmann, D.~C. Hooper, M.~Archidiacono, J.~Lesgourgues, and T.~Sprenger,
  ``{The promising future of a robust cosmological neutrino mass
  measurement},'' \href{http://dx.doi.org/10.1088/1475-7516/2019/01/059}{{\em
  JCAP} {\bfseries 01} (2019) 059},
  \href{http://arxiv.org/abs/1808.05955}{{\ttfamily arXiv:1808.05955
  [astro-ph.CO]}}.

\bibitem{Smith:2016lnt}
K.~M. Smith and S.~Ferraro, ``{Detecting Patchy Reionization in the Cosmic
  Microwave Background},''
  \href{http://dx.doi.org/10.1103/PhysRevLett.119.021301}{{\em Phys. Rev.
  Lett.} {\bfseries 119} no.~2, (2017) 021301},
  \href{http://arxiv.org/abs/1607.01769}{{\ttfamily arXiv:1607.01769
  [astro-ph.CO]}}.

\bibitem{Ferraro:2018izc}
S.~Ferraro and K.~M. Smith, ``{Characterizing the epoch of reionization with
  the small-scale CMB: Constraints on the optical depth and duration},''
  \href{http://dx.doi.org/10.1103/PhysRevD.98.123519}{{\em Phys. Rev. D}
  {\bfseries 98} no.~12, (2018) 123519},
  \href{http://arxiv.org/abs/1803.07036}{{\ttfamily arXiv:1803.07036
  [astro-ph.CO]}}.

\bibitem{Alvarez:2020gvl}
M.~A. Alvarez, S.~Ferraro, J.~C. Hill, R.~Hlo\v{z}ek, and M.~Ikape,
  ``{Mitigating the optical depth degeneracy using the kinematic
  Sunyaev-Zel'dovich effect with CMB-S4},''
  \href{http://dx.doi.org/10.1103/PhysRevD.103.063518}{{\em Phys. Rev. D}
  {\bfseries 103} no.~6, (2021) 063518},
  \href{http://arxiv.org/abs/2006.06594}{{\ttfamily arXiv:2006.06594
  [astro-ph.CO]}}.

\bibitem{Abazajian:2019eic}
K.~Abazajian {\em et~al.}, ``{CMB-S4 Science Case, Reference Design, and
  Project Plan},'' \href{http://arxiv.org/abs/1907.04473}{{\ttfamily
  arXiv:1907.04473 [astro-ph.IM]}}.

\bibitem{Vagnozzi:2018jhn}
S.~Vagnozzi, S.~Dhawan, M.~Gerbino, K.~Freese, A.~Goobar, and O.~Mena,
  ``{Constraints on the sum of the neutrino masses in dynamical dark energy
  models with $w(z) \geq -1$ are tighter than those obtained in
  $\Lambda$CDM},'' \href{http://dx.doi.org/10.1103/PhysRevD.98.083501}{{\em
  Phys. Rev. D} {\bfseries 98} no.~8, (2018) 083501},
  \href{http://arxiv.org/abs/1801.08553}{{\ttfamily arXiv:1801.08553
  [astro-ph.CO]}}.

\bibitem{Berghaus:2024kra}
K.~V. Berghaus, J.~A. Kable, and V.~Miranda, ``{Quantifying Scalar Field
  Dynamics with DESI 2024 Y1 BAO measurements},''
  \href{http://arxiv.org/abs/2404.14341}{{\ttfamily arXiv:2404.14341
  [astro-ph.CO]}}.

\bibitem{Allison:2015qca}
R.~Allison, P.~Caucal, E.~Calabrese, J.~Dunkley, and T.~Louis, ``{Towards a
  cosmological neutrino mass detection},''
  \href{http://dx.doi.org/10.1103/PhysRevD.92.123535}{{\em Phys. Rev. D}
  {\bfseries 92} no.~12, (2015) 123535},
  \href{http://arxiv.org/abs/1509.07471}{{\ttfamily arXiv:1509.07471
  [astro-ph.CO]}}.

\bibitem{Hu:2001kj}
W.~Hu and T.~Okamoto, ``{Mass reconstruction with cmb polarization},''
  \href{http://dx.doi.org/10.1086/341110}{{\em Astrophys. J.} {\bfseries 574}
  (2002) 566--574}, \href{http://arxiv.org/abs/astro-ph/0111606}{{\ttfamily
  arXiv:astro-ph/0111606}}.

\bibitem{Seljak:2003pn}
U.~Seljak and C.~M. Hirata, ``{Gravitational lensing as a contaminant of the
  gravity wave signal in CMB},''
  \href{http://dx.doi.org/10.1103/PhysRevD.69.043005}{{\em Phys. Rev. D}
  {\bfseries 69} (2004) 043005},
  \href{http://arxiv.org/abs/astro-ph/0310163}{{\ttfamily
  arXiv:astro-ph/0310163}}.

\bibitem{Green:2016cjr}
D.~Green, J.~Meyers, and A.~van Engelen, ``{CMB Delensing Beyond the B
  Modes},'' \href{http://dx.doi.org/10.1088/1475-7516/2017/12/005}{{\em JCAP}
  {\bfseries 12} (2017) 005}, \href{http://arxiv.org/abs/1609.08143}{{\ttfamily
  arXiv:1609.08143 [astro-ph.CO]}}.

\bibitem{Hotinli:2021umk}
S.~C. Hotinli, J.~Meyers, C.~Trendafilova, D.~Green, and A.~van Engelen, ``{The
  benefits of CMB delensing},''
  \href{http://dx.doi.org/10.1088/1475-7516/2022/04/020}{{\em JCAP} {\bfseries
  04} no.~04, (2022) 020}, \href{http://arxiv.org/abs/2111.15036}{{\ttfamily
  arXiv:2111.15036 [astro-ph.CO]}}.

\bibitem{vanEngelen:2013rla}
A.~van Engelen, S.~Bhattacharya, N.~Sehgal, G.~P. Holder, O.~Zahn, and
  D.~Nagai, ``{CMB Lensing Power Spectrum Biases from Galaxies and Clusters
  using High-angular Resolution Temperature Maps},''
  \href{http://dx.doi.org/10.1088/0004-637X/786/1/13}{{\em Astrophys. J.}
  {\bfseries 786} (2014) 13}, \href{http://arxiv.org/abs/1310.7023}{{\ttfamily
  arXiv:1310.7023 [astro-ph.CO]}}.

\bibitem{Aalberts:2018obr}
J.~L. Aalberts {\em et~al.}, ``{Precision constraints on radiative neutrino
  decay with CMB spectral distortion},''
  \href{http://dx.doi.org/10.1103/PhysRevD.98.023001}{{\em Phys. Rev. D}
  {\bfseries 98} (2018) 023001},
  \href{http://arxiv.org/abs/1803.00588}{{\ttfamily arXiv:1803.00588
  [astro-ph.CO]}}.

\bibitem{Barenboim:2020vrr}
G.~Barenboim, J.~Z. Chen, S.~Hannestad, I.~M. Oldengott, T.~Tram, and Y.~Y.~Y.
  Wong, ``{Invisible neutrino decay in precision cosmology},''
  \href{http://dx.doi.org/10.1088/1475-7516/2021/03/087}{{\em JCAP} {\bfseries
  03} (2021) 087}, \href{http://arxiv.org/abs/2011.01502}{{\ttfamily
  arXiv:2011.01502 [astro-ph.CO]}}.

\bibitem{Chacko:2019nej}
Z.~Chacko, A.~Dev, P.~Du, V.~Poulin, and Y.~Tsai, ``{Cosmological Limits on the
  Neutrino Mass and Lifetime},''
  \href{http://dx.doi.org/10.1007/JHEP04(2020)020}{{\em JHEP} {\bfseries 04}
  (2020) 020}, \href{http://arxiv.org/abs/1909.05275}{{\ttfamily
  arXiv:1909.05275 [hep-ph]}}.

\bibitem{Chacko:2020hmh}
Z.~Chacko, A.~Dev, P.~Du, V.~Poulin, and Y.~Tsai, ``{Determining the Neutrino
  Lifetime from Cosmology},''
  \href{http://dx.doi.org/10.1103/PhysRevD.103.043519}{{\em Phys. Rev. D}
  {\bfseries 103} no.~4, (2021) 043519},
  \href{http://arxiv.org/abs/2002.08401}{{\ttfamily arXiv:2002.08401
  [astro-ph.CO]}}.

\bibitem{Escudero:2020ped}
M.~Escudero, J.~Lopez-Pavon, N.~Rius, and S.~Sandner, ``{Relaxing Cosmological
  Neutrino Mass Bounds with Unstable Neutrinos},''
  \href{http://dx.doi.org/10.1007/JHEP12(2020)119}{{\em JHEP} {\bfseries 12}
  (2020) 119}, \href{http://arxiv.org/abs/2007.04994}{{\ttfamily
  arXiv:2007.04994 [hep-ph]}}.

\bibitem{FrancoAbellan:2021hdb}
G.~Franco~Abell\'an, Z.~Chacko, A.~Dev, P.~Du, V.~Poulin, and Y.~Tsai,
  ``{Improved cosmological constraints on the neutrino mass and lifetime},''
  \href{http://dx.doi.org/10.1007/JHEP08(2022)076}{{\em JHEP} {\bfseries 08}
  (2022) 076}, \href{http://arxiv.org/abs/2112.13862}{{\ttfamily
  arXiv:2112.13862 [hep-ph]}}.

\bibitem{Escudero:2022gez}
M.~Escudero, T.~Schwetz, and J.~Terol-Calvo, ``{A seesaw model for large
  neutrino masses in concordance with cosmology},''
  \href{http://dx.doi.org/10.1007/JHEP02(2023)142}{{\em JHEP} {\bfseries 02}
  (2023) 142}, \href{http://arxiv.org/abs/2211.01729}{{\ttfamily
  arXiv:2211.01729 [hep-ph]}}.

\bibitem{KATRIN:2021uub}
{\bfseries KATRIN} Collaboration, M.~Aker {\em et~al.}, ``{Direct neutrino-mass
  measurement with sub-electronvolt sensitivity},''
  \href{http://dx.doi.org/10.1038/s41567-021-01463-1}{{\em Nature Phys.}
  {\bfseries 18} no.~2, (2022) 160--166},
  \href{http://arxiv.org/abs/2105.08533}{{\ttfamily arXiv:2105.08533
  [hep-ex]}}.

\bibitem{Farzan:2002wx}
Y.~Farzan, ``{Bounds on the coupling of the Majoron to light neutrinos from
  supernova cooling},''
  \href{http://dx.doi.org/10.1103/PhysRevD.67.073015}{{\em Phys. Rev. D}
  {\bfseries 67} (2003) 073015},
  \href{http://arxiv.org/abs/hep-ph/0211375}{{\ttfamily arXiv:hep-ph/0211375}}.

\bibitem{Chacko:2003dt}
Z.~Chacko, L.~J. Hall, T.~Okui, and S.~J. Oliver, ``{CMB signals of neutrino
  mass generation},'' \href{http://dx.doi.org/10.1103/PhysRevD.70.085008}{{\em
  Phys. Rev. D} {\bfseries 70} (2004) 085008},
  \href{http://arxiv.org/abs/hep-ph/0312267}{{\ttfamily arXiv:hep-ph/0312267}}.

\bibitem{Friedland:2007vv}
A.~Friedland, K.~M. Zurek, and S.~Bashinsky, ``{Constraining Models of Neutrino
  Mass and Neutrino Interactions with the Planck Satellite},''
  \href{http://arxiv.org/abs/0704.3271}{{\ttfamily arXiv:0704.3271
  [astro-ph]}}.

\bibitem{Archidiacono:2013dua}
M.~Archidiacono and S.~Hannestad, ``{Updated constraints on non-standard
  neutrino interactions from Planck},''
  \href{http://dx.doi.org/10.1088/1475-7516/2014/07/046}{{\em JCAP} {\bfseries
  07} (2014) 046}, \href{http://arxiv.org/abs/1311.3873}{{\ttfamily
  arXiv:1311.3873 [astro-ph.CO]}}.

\bibitem{Baumann:2016wac}
D.~Baumann, D.~Green, and B.~Wallisch, ``{New Target for Cosmic Axion
  Searches},'' \href{http://dx.doi.org/10.1103/PhysRevLett.117.171301}{{\em
  Phys. Rev. Lett.} {\bfseries 117} no.~17, (2016) 171301},
  \href{http://arxiv.org/abs/1604.08614}{{\ttfamily arXiv:1604.08614
  [astro-ph.CO]}}.

\bibitem{Gelmini:1983ea}
G.~B. Gelmini and J.~W.~F. Valle, ``{Fast Invisible Neutrino Decays},''
  \href{http://dx.doi.org/10.1016/0370-2693(84)91258-9}{{\em Phys. Lett. B}
  {\bfseries 142} (1984) 181--187}.

\bibitem{Ekhterachian:2021rkx}
M.~Ekhterachian, A.~Hook, S.~Kumar, and Y.~Tsai, ``{Bounds on gauge bosons
  coupled to nonconserved currents},''
  \href{http://dx.doi.org/10.1103/PhysRevD.104.035034}{{\em Phys. Rev. D}
  {\bfseries 104} no.~3, (2021) 035034},
  \href{http://arxiv.org/abs/2103.13396}{{\ttfamily arXiv:2103.13396
  [hep-ph]}}.

\bibitem{Beacom:2004yd}
J.~F. Beacom, N.~F. Bell, and S.~Dodelson, ``{Neutrinoless universe},''
  \href{http://dx.doi.org/10.1103/PhysRevLett.93.121302}{{\em Phys. Rev. Lett.}
  {\bfseries 93} (2004) 121302},
  \href{http://arxiv.org/abs/astro-ph/0404585}{{\ttfamily
  arXiv:astro-ph/0404585}}.

\bibitem{Farzan:2015pca}
Y.~Farzan and S.~Hannestad, ``{Neutrinos secretly converting to lighter
  particles to please both KATRIN and the cosmos},''
  \href{http://dx.doi.org/10.1088/1475-7516/2016/02/058}{{\em JCAP} {\bfseries
  02} (2016) 058}, \href{http://arxiv.org/abs/1510.02201}{{\ttfamily
  arXiv:1510.02201 [hep-ph]}}.

\bibitem{Mangano:2005cc}
G.~Mangano, G.~Miele, S.~Pastor, T.~Pinto, O.~Pisanti, and P.~D. Serpico,
  ``{Relic neutrino decoupling including flavor oscillations},''
  \href{http://dx.doi.org/10.1016/j.nuclphysb.2005.09.041}{{\em Nucl. Phys. B}
  {\bfseries 729} (2005) 221--234},
  \href{http://arxiv.org/abs/hep-ph/0506164}{{\ttfamily arXiv:hep-ph/0506164}}.

\bibitem{EscuderoAbenza:2020cmq}
M.~Escudero~Abenza, ``{Precision early universe thermodynamics made simple:
  $N_{\rm eff}$ and neutrino decoupling in the Standard Model and beyond},''
  \href{http://dx.doi.org/10.1088/1475-7516/2020/05/048}{{\em JCAP} {\bfseries
  05} (2020) 048}, \href{http://arxiv.org/abs/2001.04466}{{\ttfamily
  arXiv:2001.04466 [hep-ph]}}.

\bibitem{Akita:2020szl}
K.~Akita and M.~Yamaguchi, ``{A precision calculation of relic neutrino
  decoupling},'' \href{http://dx.doi.org/10.1088/1475-7516/2020/08/012}{{\em
  JCAP} {\bfseries 08} (2020) 012},
  \href{http://arxiv.org/abs/2005.07047}{{\ttfamily arXiv:2005.07047
  [hep-ph]}}.

\bibitem{Froustey:2020mcq}
J.~Froustey, C.~Pitrou, and M.~C. Volpe, ``{Neutrino decoupling including
  flavour oscillations and primordial nucleosynthesis},''
  \href{http://dx.doi.org/10.1088/1475-7516/2020/12/015}{{\em JCAP} {\bfseries
  12} (2020) 015}, \href{http://arxiv.org/abs/2008.01074}{{\ttfamily
  arXiv:2008.01074 [hep-ph]}}.

\bibitem{Bennett:2020zkv}
J.~J. Bennett, G.~Buldgen, P.~F. De~Salas, M.~Drewes, S.~Gariazzo, S.~Pastor,
  and Y.~Y.~Y. Wong, ``{Towards a precision calculation of $N_{\rm eff}$ in the
  Standard Model II: Neutrino decoupling in the presence of flavour
  oscillations and finite-temperature QED},''
  \href{http://dx.doi.org/10.1088/1475-7516/2021/04/073}{{\em JCAP} {\bfseries
  04} (2021) 073}, \href{http://arxiv.org/abs/2012.02726}{{\ttfamily
  arXiv:2012.02726 [hep-ph]}}.

\bibitem{Bond:2024ivb}
J.~R. Bond, G.~M. Fuller, E.~Grohs, J.~Meyers, and M.~J. Wilson, ``{Cosmic
  Neutrino Decoupling and its Observable Imprints: Insights from Entropic-Dual
  Transport},'' \href{http://arxiv.org/abs/2403.19038}{{\ttfamily
  arXiv:2403.19038 [astro-ph.CO]}}.

\bibitem{Fields:2019pfx}
B.~D. Fields, K.~A. Olive, T.-H. Yeh, and C.~Young, ``{Big-Bang Nucleosynthesis
  after Planck},'' \href{http://dx.doi.org/10.1088/1475-7516/2020/03/010}{{\em
  JCAP} {\bfseries 03} (2020) 010},
  \href{http://arxiv.org/abs/1912.01132}{{\ttfamily arXiv:1912.01132
  [astro-ph.CO]}}. [Erratum: JCAP 11, E02 (2020)].

\bibitem{Green:2021gdc}
D.~Green, D.~E. Kaplan, and S.~Rajendran, ``{Neutrino interactions in the late
  universe},'' \href{http://dx.doi.org/10.1007/JHEP11(2021)162}{{\em JHEP}
  {\bfseries 11} (2021) 162}, \href{http://arxiv.org/abs/2108.06928}{{\ttfamily
  arXiv:2108.06928 [hep-ph]}}.

\bibitem{Nollett:2014lwa}
K.~M. Nollett and G.~Steigman, ``{BBN And The CMB Constrain Neutrino Coupled
  Light WIMPs},'' \href{http://dx.doi.org/10.1103/PhysRevD.91.083505}{{\em
  Phys. Rev. D} {\bfseries 91} no.~8, (2015) 083505},
  \href{http://arxiv.org/abs/1411.6005}{{\ttfamily arXiv:1411.6005
  [astro-ph.CO]}}.

\bibitem{SDSS:2004kjl}
{\bfseries SDSS} Collaboration, P.~McDonald {\em et~al.}, ``{The Lyman-alpha
  forest power spectrum from the Sloan Digital Sky Survey},''
  \href{http://dx.doi.org/10.1086/444361}{{\em Astrophys. J. Suppl.} {\bfseries
  163} (2006) 80--109}, \href{http://arxiv.org/abs/astro-ph/0405013}{{\ttfamily
  arXiv:astro-ph/0405013}}.

\bibitem{Viel:2005qj}
M.~Viel, J.~Lesgourgues, M.~G. Haehnelt, S.~Matarrese, and A.~Riotto,
  ``{Constraining warm dark matter candidates including sterile neutrinos and
  light gravitinos with WMAP and the Lyman-alpha forest},''
  \href{http://dx.doi.org/10.1103/PhysRevD.71.063534}{{\em Phys. Rev. D}
  {\bfseries 71} (2005) 063534},
  \href{http://arxiv.org/abs/astro-ph/0501562}{{\ttfamily
  arXiv:astro-ph/0501562}}.

\bibitem{Boyarsky:2008xj}
A.~Boyarsky, J.~Lesgourgues, O.~Ruchayskiy, and M.~Viel, ``{Lyman-alpha
  constraints on warm and on warm-plus-cold dark matter models},''
  \href{http://dx.doi.org/10.1088/1475-7516/2009/05/012}{{\em JCAP} {\bfseries
  05} (2009) 012}, \href{http://arxiv.org/abs/0812.0010}{{\ttfamily
  arXiv:0812.0010 [astro-ph]}}.

\bibitem{Xu:2018efh}
W.~L. Xu, C.~Dvorkin, and A.~Chael, ``{Probing sub-GeV Dark Matter-Baryon
  Scattering with Cosmological Observables},''
  \href{http://dx.doi.org/10.1103/PhysRevD.97.103530}{{\em Phys. Rev. D}
  {\bfseries 97} no.~10, (2018) 103530},
  \href{http://arxiv.org/abs/1802.06788}{{\ttfamily arXiv:1802.06788
  [astro-ph.CO]}}.

\bibitem{Nadler:2019zrb}
E.~O. Nadler, V.~Gluscevic, K.~K. Boddy, and R.~H. Wechsler, ``{Constraints on
  Dark Matter Microphysics from the Milky Way Satellite Population},''
  \href{http://dx.doi.org/10.3847/2041-8213/ab1eb2}{{\em Astrophys. J. Lett.}
  {\bfseries 878} no.~2, (2019) 32},
  \href{http://arxiv.org/abs/1904.10000}{{\ttfamily arXiv:1904.10000
  [astro-ph.CO]}}. [Erratum: Astrophys.J.Lett. 897, L46 (2020), Erratum:
  Astrophys.J. 897, L46 (2020)].

\bibitem{Maamari:2020aqz}
K.~Maamari, V.~Gluscevic, K.~K. Boddy, E.~O. Nadler, and R.~H. Wechsler,
  ``{Bounds on velocity-dependent dark matter-proton scattering from Milky Way
  satellite abundance},''
  \href{http://dx.doi.org/10.3847/2041-8213/abd807}{{\em Astrophys. J. Lett.}
  {\bfseries 907} no.~2, (2021) L46},
  \href{http://arxiv.org/abs/2010.02936}{{\ttfamily arXiv:2010.02936
  [astro-ph.CO]}}.

\bibitem{Kofman:1986am}
L.~Kofman, D.~Pogosyan, and A.~A. Starobinsky, ``{The large scale microwave
  backbround anisotropy in unstable cosmologies},'' {\em Sov. Astron. Lett.}
  {\bfseries 12} (1986) 175--179.

\bibitem{DeLopeAmigo:2009dc}
S.~De~Lope~Amigo, W.~M.-Y. Cheung, Z.~Huang, and S.-P. Ng, ``{Cosmological
  Constraints on Decaying Dark Matter},''
  \href{http://dx.doi.org/10.1088/1475-7516/2009/06/005}{{\em JCAP} {\bfseries
  06} (2009) 005}, \href{http://arxiv.org/abs/0812.4016}{{\ttfamily
  arXiv:0812.4016 [hep-ph]}}.

\bibitem{Audren:2014bca}
B.~Audren, J.~Lesgourgues, G.~Mangano, P.~D. Serpico, and T.~Tram, ``{Strongest
  model-independent bound on the lifetime of Dark Matter},''
  \href{http://dx.doi.org/10.1088/1475-7516/2014/12/028}{{\em JCAP} {\bfseries
  12} (2014) 028}, \href{http://arxiv.org/abs/1407.2418}{{\ttfamily
  arXiv:1407.2418 [astro-ph.CO]}}.

\bibitem{Poulin:2016nat}
V.~Poulin, P.~D. Serpico, and J.~Lesgourgues, ``{A fresh look at linear
  cosmological constraints on a decaying dark matter component},''
  \href{http://dx.doi.org/10.1088/1475-7516/2016/08/036}{{\em JCAP} {\bfseries
  08} (2016) 036}, \href{http://arxiv.org/abs/1606.02073}{{\ttfamily
  arXiv:1606.02073 [astro-ph.CO]}}.

\bibitem{Fardon:2003eh}
R.~Fardon, A.~E. Nelson, and N.~Weiner, ``{Dark energy from mass varying
  neutrinos},'' \href{http://dx.doi.org/10.1088/1475-7516/2004/10/005}{{\em
  JCAP} {\bfseries 10} (2004) 005},
  \href{http://arxiv.org/abs/astro-ph/0309800}{{\ttfamily
  arXiv:astro-ph/0309800}}.

\bibitem{Lorenz:2018fzb}
C.~S. Lorenz, L.~Funcke, E.~Calabrese, and S.~Hannestad, ``{Time-varying
  neutrino mass from a supercooled phase transition: current cosmological
  constraints and impact on the $\Omega_m$-$\sigma_8$ plane},''
  \href{http://dx.doi.org/10.1103/PhysRevD.99.023501}{{\em Phys. Rev. D}
  {\bfseries 99} no.~2, (2019) 023501},
  \href{http://arxiv.org/abs/1811.01991}{{\ttfamily arXiv:1811.01991
  [astro-ph.CO]}}.

\bibitem{Lorenz:2021alz}
C.~S. Lorenz, L.~Funcke, M.~L\"offler, and E.~Calabrese, ``{Reconstruction of
  the neutrino mass as a function of redshift},''
  \href{http://dx.doi.org/10.1103/PhysRevD.104.123518}{{\em Phys. Rev. D}
  {\bfseries 104} no.~12, (2021) 123518},
  \href{http://arxiv.org/abs/2102.13618}{{\ttfamily arXiv:2102.13618
  [astro-ph.CO]}}.

\bibitem{Davoudiasl:2018hjw}
H.~Davoudiasl, G.~Mohlabeng, and M.~Sullivan, ``{Galactic Dark Matter
  Population as the Source of Neutrino Masses},''
  \href{http://dx.doi.org/10.1103/PhysRevD.98.021301}{{\em Phys. Rev. D}
  {\bfseries 98} no.~2, (2018) 021301},
  \href{http://arxiv.org/abs/1803.00012}{{\ttfamily arXiv:1803.00012
  [hep-ph]}}.

\bibitem{Cyr-Racine:2021oal}
F.-Y. Cyr-Racine, F.~Ge, and L.~Knox, ``{Symmetry of Cosmological Observables,
  a Mirror World Dark Sector, and the Hubble Constant},''
  \href{http://dx.doi.org/10.1103/PhysRevLett.128.201301}{{\em Phys. Rev.
  Lett.} {\bfseries 128} no.~20, (2022) 201301},
  \href{http://arxiv.org/abs/2107.13000}{{\ttfamily arXiv:2107.13000
  [astro-ph.CO]}}.

\bibitem{Ge:2022qws}
F.~Ge, F.-Y. Cyr-Racine, and L.~Knox, ``{Scaling transformations and the
  origins of light relics constraints from cosmic microwave background
  observations},'' \href{http://dx.doi.org/10.1103/PhysRevD.107.023517}{{\em
  Phys. Rev. D} {\bfseries 107} no.~2, (2023) 023517},
  \href{http://arxiv.org/abs/2210.16335}{{\ttfamily arXiv:2210.16335
  [astro-ph.CO]}}.

\bibitem{Arkani-Hamed:2016rle}
N.~Arkani-Hamed, T.~Cohen, R.~T. D'Agnolo, A.~Hook, H.~D. Kim, and D.~Pinner,
  ``{Solving the Hierarchy Problem at Reheating with a Large Number of Degrees
  of Freedom},'' \href{http://dx.doi.org/10.1103/PhysRevLett.117.251801}{{\em
  Phys. Rev. Lett.} {\bfseries 117} no.~25, (2016) 251801},
  \href{http://arxiv.org/abs/1607.06821}{{\ttfamily arXiv:1607.06821
  [hep-ph]}}.

\bibitem{Chacko:2016hvu}
Z.~Chacko, N.~Craig, P.~J. Fox, and R.~Harnik, ``{Cosmology in Mirror Twin
  Higgs and Neutrino Masses},''
  \href{http://dx.doi.org/10.1007/JHEP07(2017)023}{{\em JHEP} {\bfseries 07}
  (2017) 023}, \href{http://arxiv.org/abs/1611.07975}{{\ttfamily
  arXiv:1611.07975 [hep-ph]}}.

\bibitem{Chacko:2018vss}
Z.~Chacko, D.~Curtin, M.~Geller, and Y.~Tsai, ``{Cosmological Signatures of a
  Mirror Twin Higgs},'' \href{http://dx.doi.org/10.1007/JHEP09(2018)163}{{\em
  JHEP} {\bfseries 09} (2018) 163},
  \href{http://arxiv.org/abs/1803.03263}{{\ttfamily arXiv:1803.03263
  [hep-ph]}}.

\bibitem{Will:2014kxa}
C.~M. Will, ``{The Confrontation between General Relativity and Experiment},''
  \href{http://dx.doi.org/10.12942/lrr-2014-4}{{\em Living Rev. Rel.}
  {\bfseries 17} (2014) 4}, \href{http://arxiv.org/abs/1403.7377}{{\ttfamily
  arXiv:1403.7377 [gr-qc]}}.

\bibitem{Kesden:2006zb}
M.~Kesden and M.~Kamionkowski, ``{Galilean Equivalence for Galactic Dark
  Matter},'' \href{http://dx.doi.org/10.1103/PhysRevLett.97.131303}{{\em Phys.
  Rev. Lett.} {\bfseries 97} (2006) 131303},
  \href{http://arxiv.org/abs/astro-ph/0606566}{{\ttfamily
  arXiv:astro-ph/0606566}}.

\bibitem{Kesden:2006vz}
M.~Kesden and M.~Kamionkowski, ``{Tidal Tails Test the Equivalence Principle in
  the Dark Sector},'' \href{http://dx.doi.org/10.1103/PhysRevD.74.083007}{{\em
  Phys. Rev. D} {\bfseries 74} (2006) 083007},
  \href{http://arxiv.org/abs/astro-ph/0608095}{{\ttfamily
  arXiv:astro-ph/0608095}}.

\bibitem{Keselman:2009nx}
J.~A. Keselman, A.~Nusser, and P.~J.~E. Peebles, ``{Cosmology with Equivalence
  Principle Breaking in the Dark Sector},''
  \href{http://dx.doi.org/10.1103/PhysRevD.81.063521}{{\em Phys. Rev. D}
  {\bfseries 81} (2010) 063521},
  \href{http://arxiv.org/abs/0912.4177}{{\ttfamily arXiv:0912.4177
  [astro-ph.CO]}}.

\bibitem{Bogorad:2023wzn}
Z.~Bogorad, P.~W. Graham, and H.~Ramani, ``{Coherent Self-Interactions of Dark
  Matter in the Bullet Cluster},''
  \href{http://arxiv.org/abs/2311.07648}{{\ttfamily arXiv:2311.07648
  [hep-ph]}}.

\bibitem{Archidiacono:2022iuu}
M.~Archidiacono, E.~Castorina, D.~Redigolo, and E.~Salvioni, ``{Unveiling dark
  fifth forces with linear cosmology},''
  \href{http://dx.doi.org/10.1088/1475-7516/2022/10/074}{{\em JCAP} {\bfseries
  10} (2022) 074}, \href{http://arxiv.org/abs/2204.08484}{{\ttfamily
  arXiv:2204.08484 [astro-ph.CO]}}.

\bibitem{Bottaro:2023wkd}
S.~Bottaro, E.~Castorina, M.~Costa, D.~Redigolo, and E.~Salvioni, ``{Unveiling
  dark forces with the Large Scale Structure of the Universe},''
  \href{http://arxiv.org/abs/2309.11496}{{\ttfamily arXiv:2309.11496
  [astro-ph.CO]}}.

\bibitem{Creminelli:2013mca}
P.~Creminelli, J.~Nore\~na, M.~Simonovi\'c, and F.~Vernizzi, ``{Single-Field
  Consistency Relations of Large Scale Structure},''
  \href{http://dx.doi.org/10.1088/1475-7516/2013/12/025}{{\em JCAP} {\bfseries
  12} (2013) 025}, \href{http://arxiv.org/abs/1309.3557}{{\ttfamily
  arXiv:1309.3557 [astro-ph.CO]}}.

\bibitem{Creminelli:2013poa}
P.~Creminelli, J.~Gleyzes, M.~Simonovi\'c, and F.~Vernizzi, ``{Single-Field
  Consistency Relations of Large Scale Structure. Part II: Resummation and
  Redshift Space},''
  \href{http://dx.doi.org/10.1088/1475-7516/2014/02/051}{{\em JCAP} {\bfseries
  02} (2014) 051}, \href{http://arxiv.org/abs/1311.0290}{{\ttfamily
  arXiv:1311.0290 [astro-ph.CO]}}.

\bibitem{Creminelli:2013nua}
P.~Creminelli, J.~Gleyzes, L.~Hui, M.~Simonovi\'c, and F.~Vernizzi,
  ``{Single-Field Consistency Relations of Large Scale Structure. Part III:
  Test of the Equivalence Principle},''
  \href{http://dx.doi.org/10.1088/1475-7516/2014/06/009}{{\em JCAP} {\bfseries
  06} (2014) 009}, \href{http://arxiv.org/abs/1312.6074}{{\ttfamily
  arXiv:1312.6074 [astro-ph.CO]}}.

\bibitem{Saracco:2009df}
F.~Saracco, M.~Pietroni, N.~Tetradis, V.~Pettorino, and G.~Robbers,
  ``{Non-linear Matter Spectra in Coupled Quintessence},''
  \href{http://dx.doi.org/10.1103/PhysRevD.82.023528}{{\em Phys. Rev. D}
  {\bfseries 82} (2010) 023528},
  \href{http://arxiv.org/abs/0911.5396}{{\ttfamily arXiv:0911.5396
  [astro-ph.CO]}}.

\bibitem{Bai:2023lyf}
Y.~Bai, S.~Lu, and N.~Orlofsky, ``{Gravitational Waves From More Attractive
  Dark Binaries},'' \href{http://arxiv.org/abs/2312.13378}{{\ttfamily
  arXiv:2312.13378 [astro-ph.CO]}}.

\bibitem{Hu:2001fa}
W.~Hu, ``{Angular trispectrum of the CMB},''
  \href{http://dx.doi.org/10.1103/PhysRevD.64.083005}{{\em Phys. Rev. D}
  {\bfseries 64} (2001) 083005},
  \href{http://arxiv.org/abs/astro-ph/0105117}{{\ttfamily
  arXiv:astro-ph/0105117}}.

\bibitem{Okamoto:2002ik}
T.~Okamoto and W.~Hu, ``{The angular trispectra of CMB temperature and
  polarization},'' \href{http://dx.doi.org/10.1103/PhysRevD.66.063008}{{\em
  Phys. Rev. D} {\bfseries 66} (2002) 063008},
  \href{http://arxiv.org/abs/astro-ph/0206155}{{\ttfamily
  arXiv:astro-ph/0206155}}.

\bibitem{Hanson:2009gu}
D.~Hanson and A.~Lewis, ``{Estimators for CMB Statistical Anisotropy},''
  \href{http://dx.doi.org/10.1103/PhysRevD.80.063004}{{\em Phys. Rev. D}
  {\bfseries 80} (2009) 063004},
  \href{http://arxiv.org/abs/0908.0963}{{\ttfamily arXiv:0908.0963
  [astro-ph.CO]}}.

\bibitem{Marzouk:2022utf}
K.~Marzouk, A.~Lewis, and J.~Carron, ``{Constraints on $\tau_{NL}$ from Planck
  temperature and polarization},''
  \href{http://dx.doi.org/10.1088/1475-7516/2022/08/015}{{\em JCAP} {\bfseries
  08} no.~08, (2022) 015}, \href{http://arxiv.org/abs/2205.14408}{{\ttfamily
  arXiv:2205.14408 [astro-ph.CO]}}.

\bibitem{Smith:2015uia}
K.~M. Smith, L.~Senatore, and M.~Zaldarriaga, ``{Optimal analysis of the CMB
  trispectrum},'' \href{http://arxiv.org/abs/1502.00635}{{\ttfamily
  arXiv:1502.00635 [astro-ph.CO]}}.

\bibitem{Smith:2011if}
K.~M. Smith, M.~LoVerde, and M.~Zaldarriaga, ``{A universal bound on N-point
  correlations from inflation},''
  \href{http://dx.doi.org/10.1103/PhysRevLett.107.191301}{{\em Phys. Rev.
  Lett.} {\bfseries 107} (2011) 191301},
  \href{http://arxiv.org/abs/1108.1805}{{\ttfamily arXiv:1108.1805
  [astro-ph.CO]}}.

\bibitem{Green:2023ids}
D.~Green, Y.~Huang, C.-H. Shen, and D.~Baumann, ``{Positivity from Cosmological
  Correlators},'' \href{http://dx.doi.org/10.1007/JHEP04(2024)034}{{\em JHEP}
  {\bfseries 04} (2024) 034}, \href{http://arxiv.org/abs/2310.02490}{{\ttfamily
  arXiv:2310.02490 [hep-th]}}.

\bibitem{Planck:2019kim}
{\bfseries Planck} Collaboration, Y.~Akrami {\em et~al.}, ``{Planck 2018
  results. IX. Constraints on primordial non-Gaussianity},''
  \href{http://dx.doi.org/10.1051/0004-6361/201935891}{{\em Astron. Astrophys.}
  {\bfseries 641} (2020) A9}, \href{http://arxiv.org/abs/1905.05697}{{\ttfamily
  arXiv:1905.05697 [astro-ph.CO]}}.

\bibitem{Ferraro:2014jba}
S.~Ferraro and K.~M. Smith, ``{Using large scale structure to measure $f_{NL},
  g_{NL}$ and $\tau_{NL}$},''
  \href{http://dx.doi.org/10.1103/PhysRevD.91.043506}{{\em Phys. Rev. D}
  {\bfseries 91} no.~4, (2015) 043506},
  \href{http://arxiv.org/abs/1408.3126}{{\ttfamily arXiv:1408.3126
  [astro-ph.CO]}}.

\bibitem{Gong:2011gx}
J.-O. Gong and S.~Yokoyama, ``{Scale dependent bias from primordial
  non-Gaussianity with trispectrum},''
  \href{http://dx.doi.org/10.1111/j.1745-3933.2011.01124.x}{{\em Mon. Not. Roy.
  Astron. Soc.} {\bfseries 417} (2011) 79},
  \href{http://arxiv.org/abs/1106.4404}{{\ttfamily arXiv:1106.4404
  [astro-ph.CO]}}.

\bibitem{Baumann:2012bc}
D.~Baumann, S.~Ferraro, D.~Green, and K.~M. Smith, ``{Stochastic Bias from
  Non-Gaussian Initial Conditions},''
  \href{http://dx.doi.org/10.1088/1475-7516/2013/05/001}{{\em JCAP} {\bfseries
  05} (2013) 001}, \href{http://arxiv.org/abs/1209.2173}{{\ttfamily
  arXiv:1209.2173 [astro-ph.CO]}}.

\bibitem{AnilKumar:2022flx}
N.~Anil~Kumar, G.~Sato-Polito, M.~Kamionkowski, and S.~C. Hotinli,
  ``{Primordial trispectrum from kinetic Sunyaev-Zel\textquoteright{}dovich
  tomography},'' \href{http://dx.doi.org/10.1103/PhysRevD.106.063533}{{\em
  Phys. Rev. D} {\bfseries 106} no.~6, (2022) 063533},
  \href{http://arxiv.org/abs/2205.03423}{{\ttfamily arXiv:2205.03423
  [astro-ph.CO]}}.

\bibitem{Planck:2018lbu}
{\bfseries Planck} Collaboration, N.~Aghanim {\em et~al.}, ``{Planck 2018
  results. VIII. Gravitational lensing},''
  \href{http://dx.doi.org/10.1051/0004-6361/201833886}{{\em Astron. Astrophys.}
  {\bfseries 641} (2020) A8}, \href{http://arxiv.org/abs/1807.06210}{{\ttfamily
  arXiv:1807.06210 [astro-ph.CO]}}.

\bibitem{Green:2023uyz}
D.~Green, Y.~Guo, J.~Han, and B.~Wallisch, ``{Light Fields during Inflation
  from BOSS and Future Galaxy Surveys},''
  \href{http://arxiv.org/abs/2311.04882}{{\ttfamily arXiv:2311.04882
  [astro-ph.CO]}}.

\bibitem{EUCLID:2011zbd}
{\bfseries EUCLID} Collaboration, R.~Laureijs {\em et~al.}, ``{Euclid
  Definition Study Report},'' \href{http://arxiv.org/abs/1110.3193}{{\ttfamily
  arXiv:1110.3193 [astro-ph.CO]}}.

\bibitem{SPHEREx:2014bgr}
{\bfseries SPHEREx} Collaboration, O.~Dor\'e {\em et~al.}, ``{Cosmology with
  the SPHEREX All-Sky Spectral Survey},''
  \href{http://arxiv.org/abs/1412.4872}{{\ttfamily arXiv:1412.4872
  [astro-ph.CO]}}.

\bibitem{Shiveshwarkar:2023afl}
C.~Shiveshwarkar, T.~Brinckmann, and M.~Loverde, ``{Constraining general
  multi-field inflation using the SPHEREx all-sky survey},''
  \href{http://arxiv.org/abs/2312.15038}{{\ttfamily arXiv:2312.15038
  [astro-ph.CO]}}.

\bibitem{Green:2014xqa}
D.~Green, ``{Disorder in the Early Universe},''
  \href{http://dx.doi.org/10.1088/1475-7516/2015/03/020}{{\em JCAP} {\bfseries
  03} (2015) 020}, \href{http://arxiv.org/abs/1409.6698}{{\ttfamily
  arXiv:1409.6698 [hep-th]}}.

\bibitem{Forconi:2023akg}
M.~Forconi, E.~Di~Valentino, A.~Melchiorri, and S.~Pan, ``{A Possible Impact of
  Non-Gaussianities on Cosmological Constraints in Neutrino Physics},''
  \href{http://arxiv.org/abs/2311.04038}{{\ttfamily arXiv:2311.04038
  [astro-ph.CO]}}.

\bibitem{Takada:2013wfa}
M.~Takada and W.~Hu, ``{Power Spectrum Super-Sample Covariance},''
  \href{http://dx.doi.org/10.1103/PhysRevD.87.123504}{{\em Phys. Rev. D}
  {\bfseries 87} no.~12, (2013) 123504},
  \href{http://arxiv.org/abs/1302.6994}{{\ttfamily arXiv:1302.6994
  [astro-ph.CO]}}.

\bibitem{Xu:2021rwg}
W.~L. Xu, J.~B. Mu\~noz, and C.~Dvorkin, ``{Cosmological constraints on light
  but massive relics},''
  \href{http://dx.doi.org/10.1103/PhysRevD.105.095029}{{\em Phys. Rev. D}
  {\bfseries 105} no.~9, (2022) 095029},
  \href{http://arxiv.org/abs/2107.09664}{{\ttfamily arXiv:2107.09664
  [astro-ph.CO]}}.

\bibitem{Moroi:1993mb}
T.~Moroi, H.~Murayama, and M.~Yamaguchi, ``{Cosmological constraints on the
  light stable gravitino},''
  \href{http://dx.doi.org/10.1016/0370-2693(93)91434-O}{{\em Phys. Lett. B}
  {\bfseries 303} (1993) 289--294}.

\bibitem{Osato:2016ixc}
K.~Osato, T.~Sekiguchi, M.~Shirasaki, A.~Kamada, and N.~Yoshida,
  ``{Cosmological Constraint on the Light Gravitino Mass from CMB Lensing and
  Cosmic Shear},'' \href{http://dx.doi.org/10.1088/1475-7516/2016/06/004}{{\em
  JCAP} {\bfseries 06} (2016) 004},
  \href{http://arxiv.org/abs/1601.07386}{{\ttfamily arXiv:1601.07386
  [astro-ph.CO]}}.

\bibitem{Perez:2007ipy}
F.~P\'{e}rez and B.~Granger, ``{IPython: A System for Interactive Scientific
  Computing},'' \href{http://dx.doi.org/10.1109/MCSE.2007.53}{{\em Comput. Sci.
  Eng.} {\bfseries 9} (2007) 21}.

\bibitem{Hunter:2007mat}
J.~Hunter, ``{Matplotlib: A 2D Graphics Environment},''
  \href{http://dx.doi.org/10.1109/MCSE.2007.55}{{\em Comput. Sci. Eng.}
  {\bfseries 9} (2007) 90}.

\bibitem{Harris:2020xlr}
C.~Harris {\em et~al.}, ``{Array Programming with NumPy},''
  \href{http://dx.doi.org/10.1038/s41586-020-2649-2}{{\em Nature} {\bfseries
  585} (2020) 3572}, \href{http://arxiv.org/abs/2006.10256}{{\ttfamily
  arXiv:2006.10256 [cs.MS]}}.

\bibitem{Virtanen:2019joe}
P.~Virtanen {\em et~al.}, ``{SciPy 1.0 -- Fundamental Algorithms for Scientific
  Computing in Python},''
  \href{http://dx.doi.org/10.1038/s41592-019-0686-2}{{\em Nat. Methods}
  {\bfseries 17} (2020) 261}, \href{http://arxiv.org/abs/1907.10121}{{\ttfamily
  arXiv:1907.10121 [cs.MS]}}.

\end{thebibliography}\endgroup

\end{document}